\documentclass[onecolumn, showpacs,preprintnumbers,prb,amsmath,amssymb,superscriptaddress]{revtex4}
\makeatletter
\def\@dotsep{4,5}
\makeatother

\usepackage{graphicx}
\usepackage{bm}
\usepackage{subfigure}
\usepackage{epsfig}
\usepackage{setspace}

\newcommand{\be}{\begin{equation}}
\newcommand{\ee}{\end{equation}}
\newcommand{\ba}{\begin{eqnarray}}
\newcommand{\ea}{\end{eqnarray}}

\newcommand{\fm}{\langle f \rangle}

\newcommand{\omegaBP}{\omega_{\scriptscriptstyle\rm BP}}

\begin{document} 
\begin{spacing}{1.8}

\author{Carolina Brito}
\affiliation{Instituto de F\'{\i}sica,  Universidade Federal do Rio
        Grande do Sul, Porto Alegre, Brazil}

\affiliation{CEA -- Service de Physique de l'\'Etat
Condens\'e,~CEA~Saclay,~91191~Gif-sur-Yvette,~France}

\author{Matthieu Wyart}

\affiliation{Division of Engineering and Applied Sciences, Harvard University,
  Pierce Hall, 29 Oxford Street, Cambridge, Massachusetts 02198, USA}
\affiliation{Janelia Farm, HHMI, 19700 Helix Drive, Ashburn, VA 20147}

\date{\today}

\title{Geometric interpretation of pre-vitrification in hard sphere liquids}

\begin{abstract}

We derive a microscopic criterion for the stability of hard sphere configurations, and we show empirically that this criterion is marginally satisfied in the glass. 
This observation supports a geometric interpretation for the initial rapid rise of viscosity with packing fraction, or pre-vitrification. It also implies that barely stable 
soft modes characterize the glass structure, whose spatial extension is estimated. We show that both the short-term dynamics and activation processes occur mostly 
along those soft modes, and we study some implications of these
observations. This article synthesizes new and  previous results
[C. Brito and M. Wyart,  Euro. Phys. Letters, {\bf 76},   149-155,
(2006)  and C. Brito and M. Wyart, J. Stat. Mech., L08003
(2007) ] in a unified view.
\end{abstract}

\pacs{find packs}

\maketitle
\section{Introduction}
 
 Unlike crystals, amorphous structures are poorly understood on small length scales. 
 This is apparent when one considers the low-temperature properties of glasses such as 
heat transport \cite{andy} and the nature of the two-level systems
leading to a linear specific heat \cite{philips}, or the statistics of force chains and stress propagation in a pile of sand \cite{force}. 
Part of the difficulty comes from the out-of-equilibrium nature of amorphous solids: 
to understand their structure and properties, one must also understand how they are made. This is the difficult 
problem of the glass or jamming transition, where a fluid stops flowing and rest in some meta-stable configuration. At the center of this phenomenon lies a geometrical question: 
by which processes can a dense assembly of particles rearrange, and how do these rearrangements depend on the particles packing?

 It is surprising that a similar question has been solved in the 70's
on the apparently more complicated problem of polymers entanglement
\cite{reptation}, where the objects considered are not simple
particles but long chains forming a melt. In our view part of the
reason for this paradox is the following: in a melt, the relaxation time scales with
the length of the polymers. This fact can be captured experimentally
and is a stringent test for theories. The situation is very different
in glasses, where the length scales at play appear to be limited
\cite{heterog}. This fact makes it harder to distinguish and compare the predictions of different theories. Nevertheless, recent numerics suggests that the length scales at play may not always be small. 
Particles interacting with a purely repulsive short-range potential
display a critical point, corresponding to
jammed packings for which the overlaps between particles vanish.
Near that point, scaling laws characterize the microscopic
structure \cite{J,these_matt, torquato},  elastic 
\cite{J, these_matt, sil, matthieu3, saarloos}
and transport \cite{ning} properties, and relaxation in shear flows \cite{olsson}.  

A particularly  interesting observation is that soft modes, collective displacement
of particles with a  small restoring force, are abundant near this critical point \cite{J}. 
 The relation between the microscopic structure and the characteristic frequency and length scale of these modes was derived, and in particular the latter was shown to diverge near threshold \cite{matthieu1}. 
In turn, imposing the stability of these
modes led to the derivation of a non-trivial microscopic criterion for packing of repulsive particles \cite{matthieu2}, that any mechanically stable configuration must satisfy. 
For infinitely fast quench followed by adiabatic decompression, it was observed that this criterion is marginally satisfied \cite{J,matthieu2}: configurations generated by such a protocol  are barely stable.
This supported that at least for an infinitely fast quench,  the realization of this microscopic  criterion affects the dynamics, and suggested that soft 
modes may play a role in the structural rearrangements of particles. To show that this is  the case in the empirically relevant situation of a slow quench,
one would have to study a super-cooled liquid at finite temperature, and analyze soft modes, microscopic structure and relaxation together.
This is what we perform here, using hard spheres, where interactions are purely entropic and where a critical point  also turns out to be present, allowing a scaling analysis.
 
The paper is organized as follows.   We start by illustrating the key results on the soft modes and the stability of packing of elastic particles using a simple  model, the square lattice. In Section III, after defining the coordination of a hard sphere configuration, we establish a mapping between the free energy of a hard sphere system and the energy of an elastic network. This enables to apply all the conceptual tools developed in elastic systems to hard particles, in particular we derive a microscopic criterion for the stability of hard sphere configurations. In Section IV we present the numerical protocol we use, both in the glass and the super-cooled liquid, to identify meta-stable states and characterize their structural properties. In Section V we show that in those meta-stable states the stability criterion is saturated: configurations visited are barely stable mechanically. We confirm this observation in Section VI where the short term dynamics is studied. The marginal stability of the glass implies in particular an anomalous scaling for the mean square displacement near maximum packing, which we check numerically. In Section VII  it is shown that only a small fraction of the degrees of freedom of the system participate to activation events where new meta-stable states are visited. Those degrees of freedom are precisely the soft modes present in the glass structure.  Finally we argue that these observations support a geometric interpretation for pre-vitrification, which is presented in  Section VIII.

\section{ A criterion for the mechanical stability of elastic networks}

Studying engineering structures, Maxwell \cite{max} established a necessary criterion for the mechanical stability of elastic networks. The key microscopic parameter is the coordination $z$, the average number of interactions per particle. For an elastic network of springs, his criterion reads $z>z_c=2d$, where $d$ is the spatial dimension of the system. The demonstration goes as follows. 
Consider a set of $N$ points interacting with $N_c$ springs, at rest, of stiffness $k$. The  expansion
of the energy is: 
\be 
\label{1} 
\delta E=\sum_{\langle ij\rangle} \frac{k}{2} [(\delta {\vec
R_i}-\delta {\vec R_j})\cdot {\vec n_{ij}}]^2 +o(\delta R^2),
\ee 
where the sum is made over all springs, ${\vec n_{ij}}$ is the unit
vector going from $i$ to $j$, and $\delta {\vec R_i}$ is
the displacement of particles $i$. A system is floppy, {\it i.e.} not mechanically stable,  if it
can be deformed without energy cost, that is if there is a displacement field for which $\delta E=0$, or
equivalently $(\delta {\vec R_i}-\delta {\vec R_j})\cdot
{\vec n_{ij}}=0$ $\forall ij$. If the spatial dimension is
$d$, this linear system has $Nd$ degrees of freedom (ignoring the 
$d(d + 1)/2$ rigid motions of the entire system) and 
$N_c\equiv Nz/2$ equations, and therefore there are always non-trivial
solutions if $Nd> N_c$, that is if $z<2d\equiv z_c$. Finite stiffness therefore requires:
\be
 z\geq 2d.
\ee 

\begin{figure}
  \rotatebox{0}{\resizebox{6.0cm}{!}{\includegraphics{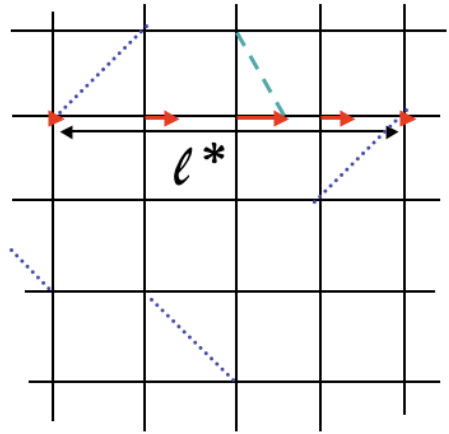}}}
  \caption{Square lattice of springs with  a density per particle
    $\delta z$ of additional diagonal springs, represented in dotted 
    lines.  $l^*\sim \sigma/\delta z$ is the typical 
    dimensionless distance of the segments contained between two 
    diagonal springs on a given row or column.  The  arrows represent the 
    longitudinal  mode of wavelength $\sim l^*$ of such a segment:
    $\delta {\vec R_i}\sim  \sin (\pi i \sigma /l^*) {\vec e_x}$, 
    following  the notation introduced in the text.  The  dashed  
    line  exemplifies the deformation of a spring 
    transverse and directly connected to the segment considered, 
    it is elongated by the longitudinal vibration of this segment.
    When the pressure is positive and contacts are under compression, 
    this elongation lowers the energy contained in those springs. This 
    leads to an elastic instability when  $\delta z$ becomes smaller 
    than a quantity proportional to the square root of the contact strain, 
    of order $f/ k \sigma$. }
  \label{f1}
\end{figure}

Under compression, the criterion of rigidity becomes more demanding. 
Here we illustrate this result in a simple model, but the different scaling we obtain have broader applications and are valid in particular for random assemblies of elastic particles \cite{matthieu1,matthieu2}. 
Consider a square lattice  of springs of rest length $\sigma$. 
It marginally satisfies the Maxwell criterion, since $z=4$ and $d=2$. 
We add randomly a density $\delta z$ of springs at rest connecting second neighbors, represented as dotted lines  in Fig(\ref{f1}), such that the coordination is $z=z_c+\delta z$.  Springs are added  in a rather homogeneous manner, 
so that there are not large regions without dotted springs. The typical distance between two dotted springs in a given row or column is then:
\be
\label{001}
 l^*\sim \sigma/\delta z
 \ee 

How much pressure $p$ can this system sustain before collapsing? 
To be mechanically stable, all  collective displacements must have a positive energetic cost. 
It turns out that the first modes to collapse as the pressure is
increased are of the type of the longitudinal  modes of wavelength
$l^*$ of individual segments of springs contained between two 
dotted diagonal springs  as represented with arrows 
in Fig(\ref{f1}).  
These modes have a displacement field of  the form $\delta {\vec R_i}=2 X \sin (\pi i  \sigma/l^*)/\sqrt{l^*/\sigma} {\vec e_x}$, where $i$ labels the particles along a segment and runs between 0 and $l^*/\sigma$, ${\vec e_x}$ is the unit vector in the direction of the line, and $X$ is the amplitude of the mode, $X=1$ for a normalized mode. 
In the absence of pressure $p$, springs carry no force. The energy  of such a mode comes only from the springs of the segment and from Eq.(\ref{1}) follows $\delta E \sim k X^2 \sigma^2/l^*{}^2 $. Note that these modes have a characteristic frequency:
\be
\label{01}
\omega^*\sim \sqrt{\frac{\delta E(X=1)}{m}}\sim \sqrt{\frac{k}{m}} \delta z
\ee
When $p>0$, each spring now carries a force of order $f\sim p
\sigma^{d-1}$. The energy expansion then contains other terms not
indicated in Eq.(\ref{1}) \cite{shlomon, matthieu2},
 whose effect can
be estimated quantitatively as follows. When particles are displaced
along a longitudinal mode such as the one represented by arrows in
Fig.(\ref{f1}), the force of each  spring directly connected and
transverse to the segment considered, represented in dashed line in
Fig(\ref{1}), now produces a work equal to $f$ times the elongation of
the spring. This elongation is simply $\delta {\vec R_i}^2/ \sigma$
following Pythagoras' theorem. Summing on all the springs transverse
to the segment leads to a work of order $ f X^2/\sigma$. This gives
finally for the energy of the mode $\delta E \sim k
X^2\sigma^2/l^*{}^2  -f X^2/\sigma$, where numerical pre-factors 
are omitted. Stability requires $\delta E > 0 $, implying that $k\sigma^3/l^*{}^2>f$, or:
\be
\label{2}
\delta z> A (f/k\sigma)^{1/2}\sim \sqrt e
\ee
where A is a numerical constant and $e$ is the typical strain in the contacts. This result signifies that  pressure has a destabilizing effect, which needs to be counterbalanced by the creation of more contacts to maintain elastic stability.
Note that Eqs.(\ref{001},\ref{01},\ref{2}) are more general that the
simple square lattice model considered here, they apply to elastic
network or assemblies of elastic particles \cite{matthieu1, matthieu2}
 as long as spatial fluctuations in coordination are limited.

\section{ An analogy between  hard sphere glasses and elastic networks }
\label{HS_elastic}

These results on the stability of elastic networks apply to hard spheres systems. In order to see that, we recall the analogy between the free energy of a hard sphere glass and the energy of an athermal network of logarithm springs  \cite{brito}. Consider the dynamics (brownian or newtonian) of hard spheres in a super-cooled liquid or glass state, such that the collision time among neighbor particles $\tau_c$ is much smaller than $\tau$, the time scale on which the structure rearranges. On  intermediate time scale $t_1$ such that $\tau_c<<t_1<<\tau$ one can define a contact network, by considering all the pairs of particles  colliding with each other, those are said to be ``in contact"  (examples of contact network are shown in the next section). This enables to define a coordination number $z$. Once the contact network is defined in a meta-stable state, all configurations for which particles in contact do not interpenetrate are equiprobable, those configurations satisfy $\prod_{\langle ij\rangle}\Theta(|| {\vec R_i}-{\vec R_j}||-\sigma)=1$, where $\Theta$ is the Heaviside function, the product is made on all contacts $ij$ and $\sigma$ is the particles diameter, that defines our unit length. The isobaric partition function is then: 
 \be
 \label{3}
 {\cal Z}=\int dV \prod_i \int d{\vec R_i} \prod_{\langle ij\rangle}\Theta(|| {\vec R_i}-{\vec R_j}||-\sigma) exp(\frac{-pV}{k_b T})
 \ee
 In one spatial dimension (for a neckless of spheres), Eq.(\ref{3}) can be readily solved by changing variables and considering the gaps $h_{ij}=R_j-R_i$ between particles in contact.
The mapping is one to one and linear:

 \be
 \label{4}
 \prod_i dR_i \propto \prod_{ij} dh_{ij}  \delta(\sum_{ij} h_{ij} - (V-V_0))
 \ee
 where $V_0$  is the volume of the system at $p=\infty$. Eqs.(\ref{3},\ref{4}) lead to:
 \be
 \label{5}
 {\cal Z}=\prod_{ij} \int_{h_{ij}\geq 0} dh_{ij} exp(\frac{-ph_{ij}}{k_b T})
\ee
leading to the simple result $p= k_b T/\langle h \rangle$. In higher dimensions, the situation is far more complicated in general, because the mapping between positions and gaps in not one-to-one, and not linear. 
 There is nevertheless an exception to that rule.  As was shown by several authors \cite{tom1, moukarzel, roux}, as the pressure diverges near maximum packing  the system becomes isostatic $z\rightarrow z_c$, see  footnote \footnote[1]{$z\geq z_c$ is imposed by the rigidity of the system. Imposing that particles do not interpenetrate and exactly touch $|| {\vec R_i}-{\vec R_j}||=\sigma$ cannot be satisfied unless $z\leq z_c$, otherwise this system is over-constrained, so that $z=z_c$ at maximum packing.} for a sketch of the argument. As noted in \cite{brito}, this implies precisely that the number of contact is equal to the number of degrees of freedom, and that the mapping of particle positions toward the gaps is one-to-one. Near maximum pressure this mapping is  also linear as $(d{\vec R_i} -d{\vec R_j})\cdot {\vec n_{ij}} =dh_{ij}+O(\delta {\vec R}^2)$. One gets:
  \be
 \label{6}
 \prod_i d{\vec R_i} \propto \prod_{ij} dh_{ij}  \delta(\sum_{ij} f_{ij} h_{ij} - p(V-V_0))
 \ee
 where $f_{ij}$ is the force in the contact $ij$. The volume constraint $ \delta(\sum_{ij} f_{ij} h_{ij} - p(V-V_0))$ generalizes for $d>1$ the constraint $\delta(\sum_{ij} h_{ij} - (V-V_0) )$.  This relation between gaps and volume can be derived as follows. In a meta-stable state, forces must be balance on all particles. As a consequence,  the virtual force theorem implies that the work of any displacement is zero: $dW=\sum_{ij} f_{ij} dh_{ij} -pdV=0$. Integrating this relation leads to the relation above . Eqs(\ref{3},\ref{6}) lead to:
 \be
\label{7}
 {\cal Z}=\prod_{ \langle ij \rangle} \int_{h_{ij}\geq 0} dh_{ij} e^{- f_{ij} h_{ij}/k_bT}.
 \ee
and:
\be
\label{10}
f_{ij}= \frac{k_b T}{ \langle h_{ij} \rangle}
\ee
The force is inversely proportional to the average gap between particles, as we shall confirm numerically in the next section. The stiffness in the contact $ij$ is then:
\be
\label{101}
k_{ij}= (k_b T)/ \langle h_{ij} \rangle^2
\ee 
From Eq.(\ref{7}) one obtains for the Gibbs free energy ${\cal G}$:
\be
\label{11}
{\cal G}= - k_b T \sum_{\langle ij\rangle} \ln(\langle h_{ij}\rangle)= - k_b T \sum_{\langle ij\rangle} \ln(r_{ij}^{eq}-\sigma)
\ee
where $r_{ij}^{eq}$ is the average distance between particle $i$ and $j$: $r_{ij}^{eq}=\langle || {\vec R_i}-{\vec R_j}||\rangle$. Thus the Gibbs free energy
of a hard sphere system is equivalent to the energy of a network of
logarithmic springs. As for an elastic network, one can define a
dynamical matrix ${\cal M}$ by differentiating Eq.(\ref{11}). ${\cal
M}$ describe the linear response of the average displacement of the
particles to any applied force field. The eigenvectors of ${\cal M}$
define the normal modes of the free energy \cite{Ashcroft}.

When $z\geq z_c$, as is the case in the glass phase (see below), Eqs.(\ref{7}-\ref{11}) are not exact. Nevertheless, the relative deviations to Eq.(\ref{10}) can be estimated \cite{these_matt},
and are of order $\delta z=z-z_c$. Numerically these corrections turn out to be small (smaller than 5\% throughout the glass phase \cite{brito}), and we shall neglect them. We will check this approximation further when we study the microscopic dynamics, see Section \ref{micrig1}. Then, together with Eq.(\ref{2}) and Eq.(\ref{101}), the present analogy leads to the prediction that minima of the free energy in hard spheres system must satisfy:
\be
\label{12}
\delta z \geq A \sqrt{e} \sim\sqrt{\langle h\rangle /\sigma}\sim  \sqrt{ \frac{k_B T}{\sigma\langle f\rangle}}.
\ee
where $\langle h\rangle$ and $\langle f\rangle$ are the typical gaps and forces between particles in contact, and where  Eq.(\ref{10}) was used to relate these two quantities.

\section{Numerical protocol}
\label{numeric}

To study if the meta-stable states visited in the super-cooled liquid and the glass live close to the bound of Eq.(\ref{12}), and if the proximity of this bound affect the dynamics, we simulate hard discs with Newtonian dynamics:  we use an event-driven code 
\cite{Allen_book}, particles are in free flight until they  collide elastically.
We use  two-dimensional bidisperse systems of $N=64, 256$ and $N=1024$ particles.
Half of the particles have a diameter $\sigma_1$, which defines our unit length. Other particles 
have diameter $\sigma_2=1.4\sigma_1$. All particles have a mass $m$, our unit mass. 
Since for hard particles $k_bT$ is only re-scaling time  and energy,  we chose $k_bT$ as our unit of energy.
Our unit of time is then $\sigma_1 \sqrt{m/k_bT}$. All data below are presented in  dimensionless quantities. 

We seek to study both the glass and the super-cooled liquid phase. To generate configurations with large packing fractions in the glass we use the
 jammed configurations  of \cite{J} with packing fraction distributed around $\phi_c\approx 0.83$.
 At $\phi_c$ the particles are in permanent contact. By reducing the particles
 diameters by a relative amount   $\epsilon$, we obtain 
 configurations of packing fraction $\phi=\phi_c(1- \epsilon)^2$.  
 We then assign a random velocity to every particle 
 and launch an event-driven simulation. 
 This procedure enables to study the aging dynamics of highly dense
 systems.  For  $\phi<\phi_0\approx 0.79$, the system is a super-cooled liquid, and can be equilibrated.

\subsection{Numerical definition of meta-stable  states}

\begin{figure}
  \rotatebox{-90}{\resizebox{6.0cm}{!}{\includegraphics{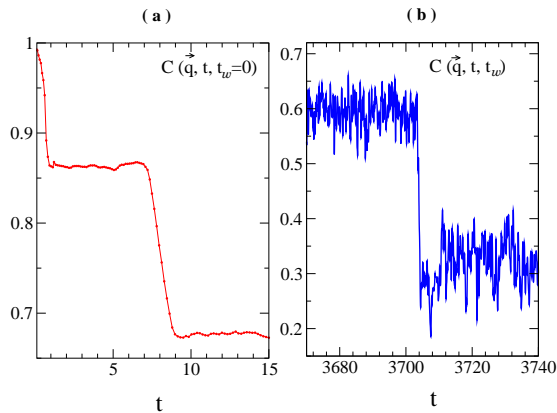}}}
  \caption{ Examples of $C(\vec q, t, t_{\omega})$ as defined in Eq.(\ref{13}) {\it vs} $t$  (a) in the glass and (b) in the     super-cooled liquid.}    
  \label{Cq_examples}
\end{figure}

Computing numerically the contact network requires time-averaging on some scale $t_1$ such that 
$\tau_c<<t_1<<\tau$, where $\tau$ is the $\alpha$-relaxation time of the system, 
which we  define  as the time for which the self scattering function decays by 70\%. 
In the super-cooled liquid a natural way to proceed would be to compute $\tau$, and chose $t_1<<\tau$.  Nevertheless this procedure is not appropriate for the aging dynamics in the glass phase, where $\tau$ is not well defined, and where the dynamics depends on the waiting time. As an alternative protocol,  
we consider the self-density  correlation function not averaged in time:
\be
\label{13}
C(\vec q, t, t_w) = \langle e^{ i \vec q . 
(\vec R_j(t+t_w) - \vec R_j(t_w) )} \rangle_j, 
\ee
where the average is made on every particle $j$ but not on time, $\vec R_j(t)$ is the  position of
particle $j$ at time $t$ and $\vec q$ is some wave vector. In what
follows  $||{\vec q}||=2\pi/\sigma_1$. For all the system sizes we consider in the glass phase, and for small systems (for $N=64$ and to a lower extent for $N=256$) near the glass transition, we observe that $C(\vec q, t, t_w)$ displays long and  well-defined plateaus interrupted by sudden jumps, as exemplified in Fig.(\ref{Cq_examples}).
In the super-cooled liquid, those jumps are of order one, indicating that the life time of the plateaus are of order $\tau$ (a few jumps de-correlate the structure). 
The existence of plateaus interrupted  by sharp transitions indicates that the dynamics is intermittent, as previously observed \cite{kob, heuer}. In real space,  the plateaus of  $C(\vec q, t, t_w)$ correspond to quiet periods where particles are rapidly rattling around their average position.
The jumps indicate rapid and collective rearrangements of the particles. In what follows we call  ``meta-stable states"  those quiet periods of the dynamics.
 Average quantities are then computed in a given meta-stable state by choosing a  time interval $[t,t+t_1]$  for which the system lies in the same meta-stable state. 
 We find that average quantities do not vary significantly with the location and the length of the time interval as long as $t_1>>\tau_c$. This robustness is proven for vibrational modes in particular in Annex 2. In what follows  we chose $t_1\sim 200 \tau_c$.  

This protocol has the advantage to be applicable both to aging and equilibrated dynamics. On the other hand,
it is limited to rather small systems in the liquid phase.  Clearly for an infinite system $C(\vec q, t, t_w)$ is spatially self-averaging, and smooth. Already for  $N=1024$ near the glass transition plateaus are hardly detectable,
and our protocol does not apply (although it does in the glass). For such system sizes, the more traditional method (computing $\tau$ from the decay of the smooth self-scattering function and considering some time scale $t_1<<\tau$)
should be used.

\subsection{Contact force network}

A straightforward quantity to define  in a meta-stable state is the average position of the particles:
\be
\label{14}
\vec R_i^{eq} = \frac{1}{t_1} \int _{t}^{t+t_1} \vec R_i(t')dt'.
\ee

\begin{figure}[htbp]
   \begin{center}
     \rotatebox{0}{\resizebox{5.0cm}{!}{\includegraphics{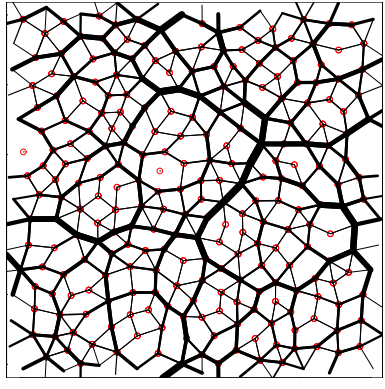}}}
     \caption{Contact forces for $N=256$, $\langle f \rangle = 6740$ and $t_1 = 10^4$ time steps. Points represent  particles centers.  Contact forces are sketched by line segments which link  particles that are in contact. The width of these segments  is proportional   to the   force amplitude. 
This figure has already been published in the reference ``On the rigidity of hard sphere glass near random close packing", Europhyscis Letters, v 76, 149-155 (2006)  by C. Brito and M. Wyart  and it  is reproduced here under permission  of the Institute of Physics Publishing (IOP). }
     \label{forcefield}
   \end{center}
\end{figure}

\begin{figure}[htbp]
   \begin{center}
     \rotatebox{-90}{\resizebox{5.5cm}{!}{\includegraphics{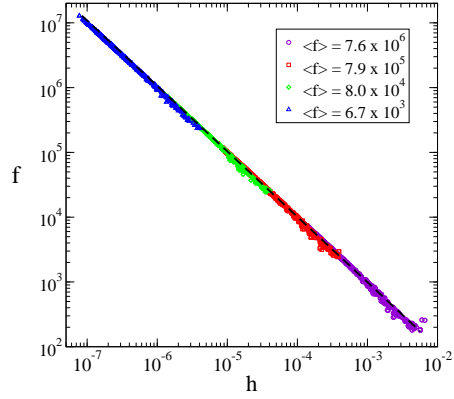}}}
     \caption{Log-log plot of the amplitude of the contact force {\it vs} the gap between the particles 
       for different values of $\langle f \rangle$ in a  system with $N=256$ particles. 
       Each point corresponds to a pair of numbers ($f_{ij}, \langle h_{ij} \rangle$) that 
       characterizes a pair of particles in contact. 
       The slashed line is a fit of the theoretical relation predicted in Eq.(\ref{10}).
       This figure has already been published in the reference ``On the rigidity of hard sphere glass near random close packing", Europhyscis Letters, v 76, 149-155 (2006)  by C. Brito and M. Wyart and it  is reproduced here under permission  of the Institute of Physics Publishing (IOP). }
     \label{f_vs_h}
   \end{center}
\end{figure}

Central to our analysis is the  definition of a contact force network
 \cite{brito, bubble, donev2}. Two   particles are said to be {\it in contact} if they collide
 with each other during the time interval $t_1$. This enables to
 define an average coordination number $z$ as $z=2N_c/N$, where $N_c$ is 
the total number of contacts among all particles of the system. 
The contact  force $\vec{f}_{ij}$ between these  particles is then 
 defined as average momentum they exchange  per unit of  time:
\be
\label{15}
\vec{f}_{ij}=\frac{1}{t_1}\sum_{n=1}^{n=n_{col}[t_1]} \Delta \vec{P}_n ,  
\ee
where the sum is made on the total number of collisions $n_{col}[t_1]$
between $i$ and $j$ that took place in the time interval $t_1$, and $\Delta
\vec{P}_n$  is the momentum exchanged at the $n$th chock. Fig.(\ref{forcefield}) shows a contact
force network obtained using this procedure, and   Fig.(\ref{f_vs_h}) shows the amplitude of the contact force 
as a function of the average gap between the particles in contact.

We define the average
 contact force of the network $\fm$ as:
\ba
 \langle f \rangle  &=&  \frac{1}{N} \sum_{\langle ij \rangle}^{N}
 || \vec f_{ij}||.
\label{16}
\ea
Near maximum packing $\fm$  scales as the pressure $p$ and as the inverse of the average gap
 $h=\langle h_{ij}\rangle$, as implied by Eq.(\ref{10}). 
The densest packing fraction $\phi_0\approx 0.79$  we can equilibrate corresponds to $\fm \approx 18$. 
For larger values of $\fm$, the system is a glass.

Note that close to maximum packing, at very large pressure, a few 
percents of the particles do not contribute to the rigidity of the structure. 
 These ``rattlers" appears in Fig.(\ref{forcefield}) as particles which 
do not exchange forces with any neighbors. In our analysis below we 
systematically remove such particles, and the procedure we use to do so is presented in Annex 1.

\subsection{Normal modes of the free energy}
\label{normalmodes}

As shown in  Eq.(\ref{11}), the free energy in a meta-stable state can be written in terms of the average particle positions. It follows that it can be expanded for small average displacements. For discs ($d=2$) this reads:
\begin{eqnarray}
\label{17}
\delta {\cal G} \approx -\sum_{\langle ij \rangle} \frac{1}{\langle h_{ij}\rangle}
\frac{[(\delta\vec{R_j}-\delta\vec{R_i})\cdot {\vec n_{ij}}^{\bot}]^2}{2 r_{ij}^{eq}  } 
+\\ \nonumber
\sum_{\langle ij \rangle} \frac{1}{2\langle h_{ij}\rangle^2} [(\delta\vec{R_j}-\delta\vec{R_i}).\vec n_{ij}]^2+o(\delta R^2)
\end{eqnarray}
where ${\vec n_{ij}}^\bot$ is the unit vector orthogonal to ${\vec n_{ij}}$. Eq.(\ref{17}) can be written in matrix form:
\be
\label{18}
\delta {\cal G} = \langle \delta {\bf R} | {\cal M} |\delta {\bf R} \rangle
\ee
where  $|\delta {\bf R} \rangle$ is the dN-dimensional vector 
$\delta\vec{R_1} ..\delta\vec{R_N}$ and 
$\langle \delta {\bf R}^\alpha|\delta {\bf R}^\beta \rangle\equiv \sum_i^N \delta\vec{R_i}^\alpha\cdot\delta\vec{R_i}^\beta$.  ${\cal M}$ is the dynamical (or stiffness) matrix. For completeness, note that for discs it
can be written as a $N \times N$ matrix whose elements 
${\cal M}_{ij}$ are tensors of rank $d$, for $d=2$ this reads:
\begin{eqnarray*}
\label{180}
{\cal M}_{ij} = -\delta _{\langle ij \rangle }
(\frac{1}{2 r_{ij}^{eq} \langle h_{ij} \rangle} \vec n_{ij}^{\bot}
\otimes \vec n_{ij}^{\bot}   -
\frac{1}{2\langle h_{ij} \rangle^2}~ \vec n_{ij} \otimes \vec n_{ij}) 
 + \\ \nonumber
\delta _{i, j }
\sum _{\langle l \rangle} 
(\frac{1}{2 r_{ij}^{eq}  \langle h_{il} \rangle}~ 
\vec n_{ij}^{\bot} \otimes \vec n_{ij}^{\bot} - 
\frac{1}{2 \langle h_{il} \rangle ^2}~ \vec n_{il} \otimes \vec n_{il}),
\end{eqnarray*}
where $\delta _{\langle ij \rangle }=1$ when particles $i$ and $j$ are in
contact and where the second sum is made on all the particles $\langle l\rangle$
 in contacts with the particle $i$. $\otimes$ is the tensor product. 
 ${\cal M}$ describes the linear response of the average displacement of the 
 particles to an external
 force.  The eingenvectors of ${\cal M}$ are the normal modes
of the system  and the  frequencies are the square  roots of these
 eigenvalues \cite{clapack}. The distribution of these frequencies is the density of states $D(\omega)$. We shall denote $|\delta {\bf R}^\omega\rangle$ the displacement 
field of a normal mode of frequency $\omega$. 
These modes form a complete orthonormal basis 
$\{|\delta {\bf R}^\omega\rangle\}$.

\section{Marginal stability of the microscopic structure}
\label{micrig}

\begin{figure}[htbp]
\begin{center} 
  \rotatebox{-90}{\resizebox{6.0cm}{!}{\includegraphics{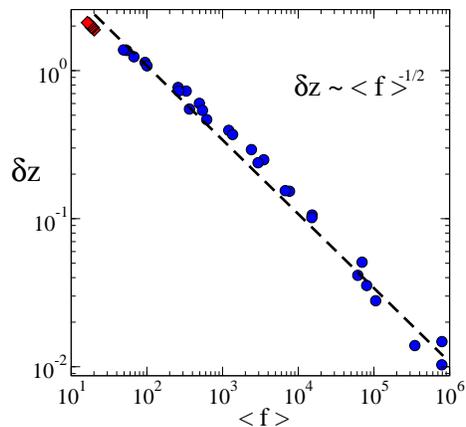}}}
    \caption{Log-log plot of $\delta z$ {\it vs} $\fm$ for $N=256$ or $N=1024$ particles. 
  Each circle correspond to one meta-stable state in the glass phase, whereas 
  diamonds correspond to averaged quantity among 11 meta-stable states  in the super-cooled liquid. 
  The slashed line is the best fit of the form $\delta z = A \fm^{-1/2}$. 
  This figure has already been published in the reference ``On the rigidity of hard sphere glass near random close packing", Europhyscis Letters, v 76, 149-155 (2006)  by C. Brito and M. Wyart and it  is reproduced here under permission  of the Institute of Physics Publishing (IOP). } 
\label{dz_vs_p}
\end{center}
\end{figure}

According to Eq.(\ref{12}), minima of the free energy must have a sufficiently coordinated contact network.
One may ask if the meta-stable states generated dynamically satisfy this bound easily, or marginally\cite{brito}. 
In order to test this question, we prepare systems at various pressures and identify
meta-stable states. Deep in the glass phase, starting from some initial condition typically 3 or 4 states are visited  during aging on the time scales we explore. During aging the pressure can drop by several orders of magnitudes (indicating the possibility to obtain denser jammed configurations under re-compression, i.e. larger $\phi_c$).
For each  meta-stable state visited, we measure the
coordination of the contact force network, and the average force $\fm$.
The corresponding data is presented in Fig.(\ref{dz_vs_p}), together with measures of the coordination in the super-cooled liquid where equilibrium is reached. 

In Fig(\ref{dz_vs_p}) it appears that the fit corresponding to the saturation of the bound of Eq.(\ref{12}), $\delta z = A \fm^{-1/2}$,  captures  well all our data-points. This observation supports that  the meta-stable states we generate lie close to {\it marginal stability}: on the time scales that can be probed numerically, the configurations visited by the dynamics have just nearly enough contacts to counter-balance the destabilizing effect  induced by the contact forces.
The situation is very different from a mono-disperse  hexagonal crystal, for which $\delta z=2$ as $\fm\rightarrow \infty$. Thus Fig(\ref{dz_vs_p}) supports that, at least for hard particles, there exists a fundamental difference in the mechanical stability of  a glass and a crystal. In what follows we provide further evidences that meta-stable states lie close to marginal stability, and study some consequences of this property on the dynamics.

\section{Microscopic dynamics}
\label{micrig1}
If a configuration is marginally rigid, then by definition it must display modes which are barely stable.
In this section we investigate the existence of such soft modes in the free energy expansion around 
meta-stable states. 
After observing that these soft modes are indeed present, we show that they lead to anomalously large
 and slow density fluctuations on time scales where the system is still confined in one meta-stable 
state, which we refer to as `` microscopic dynamics". 

\subsection{Density of States}

\begin{figure}      
  \rotatebox{-90}{\resizebox{6.0cm}{!}{\includegraphics{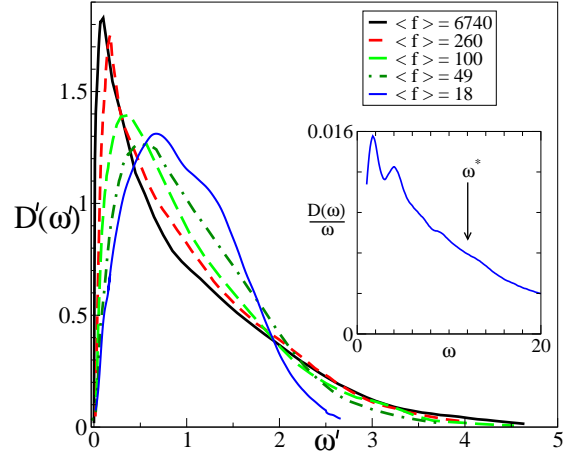}}}
  \caption{Densities of states  $D'(\omega')\equiv \fm D(\omega)$  {\it vs.} rescaled frequency
    $\omega'=\omega/\langle f\rangle$ for   different values of  $\langle f\rangle$ in a system of 
    $N=256$ particles. 
    Inset: $D(\omega)/\omega$ {\it vs.} $\omega$  for $\langle f\rangle=18$.}
  \label{Dw_examples}
\end{figure}

We compute the density of states $D(\omega)$ in meta-stables states for various
 pressure following  the procedure introduced in  section \ref{normalmodes}. As the pressure 
is varied, following Eq.(\ref{101}) the characteristic stiffness and  therefore the characteristic frequency change. It is therefore convenient to  represent the density of states in rescaled frequencies   $\omega'=\omega/\langle f\rangle$ .  
Results are shown in Fig.(\ref{Dw_examples}). Very similar results have been recently reported  in simplified ``mean field" hard sphere models \cite{kurchan}. Only the positive part of the spectrum is shown. Occasionally we observe  one or two unstable modes,  with a negative frequency, of very small absolute value.  Those unstable directions may appear due to the approximation we
 perform when computing the free energy. Alternatively, they may indicate the presence  of saddles (and multiple configurations of free 
 energy minima) or ``shoulders" in the meta-stable state under study.

\begin{figure} 
   \begin{center}
     \rotatebox{-90}{\resizebox{8.0cm}{!}{\includegraphics{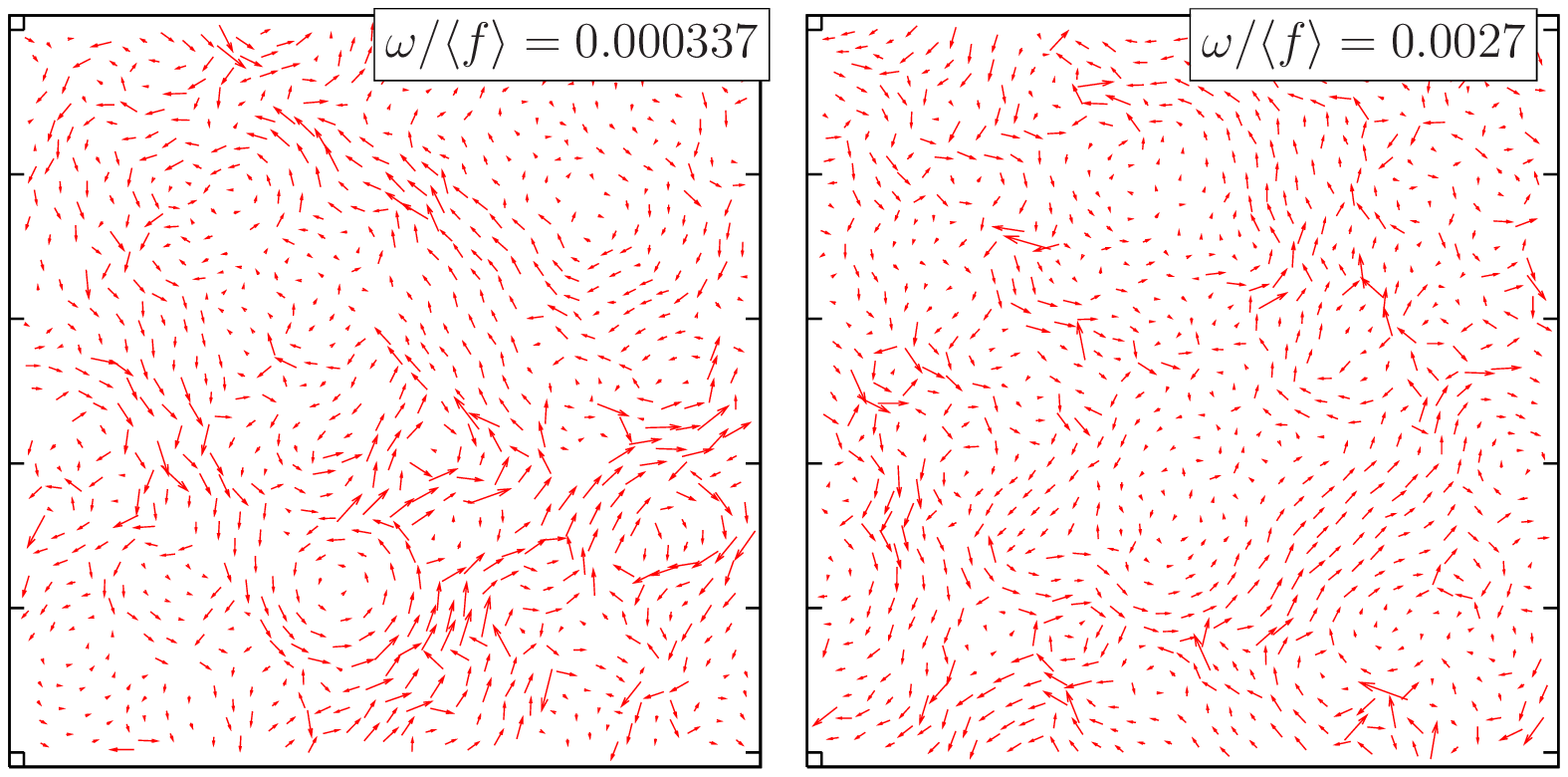}}}
     \vspace{-1.5cm}
     \caption{Examples of two lowest-frequency modes for $N=1024$ particles 
       for  $\langle f\rangle = 7.8\times 10^5$ (left) and $\langle
       f\rangle = 330$ (right).}
     \label{AVEC_examples}
     \end{center}
 \end{figure}

From Fig.(\ref{Dw_examples}) we observe that:
(i) there is an abundance of modes at low frequency. For all $\fm$, $D'(\omega ')$ 
increases rapidly from zero-frequency to reach a maximum at some 
frequency $\omega^*$, before  decaying again. In the inset of  Fig.(\ref{Dw_examples}),
 $D(\omega )$ is normalized by its Debye behavior $D_d(\omega)\sim \omega$ (plane waves would lead to a linear behavior of the density of states in 
two dimensions).  No plateau can be detected at low frequency, we rather observe a
peak  in the quantity $D(\omega)/D_d(\omega)$, which appears at some frequency 
$\omegaBP$ significantly smaller than  $\omega^*$. This indicates that for our system size we do not observe any frequency range where plane waves dominate the spectrum. This is confirmed by inspection of the lowest-frequency modes, which appear to be quite heterogeneous.  Two 
examples of  lowest-frequency modes are shown in  Fig.(\ref{AVEC_examples})
 for two values of $\fm$. Those observations are consistent with the presence of barely stable soft modes in the spectrum. 

\begin{figure}
  \rotatebox{-90}{\resizebox{6.0cm}{!}{\includegraphics{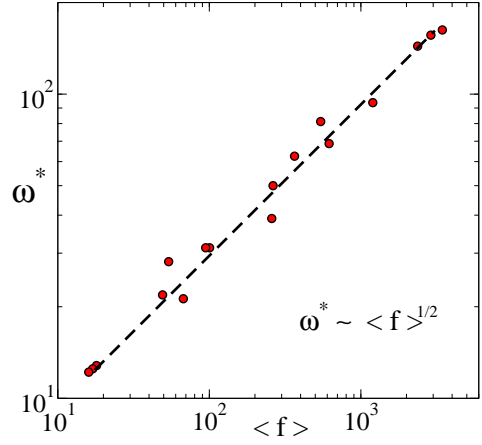}}}
   \caption{Characteristic frequency $\omega^*$ as defined in the text
   {\it vs}  average force $\fm$.}
  \label{wstar_vs_f}
\end{figure}

(ii) There exists a  characteristic frequency $\omega^*$ which scales with the pressure.
 We define $\omega^*$ as the frequency at which $D(\omega)$ is 
 maximum: $D(\omega^*)=D_{max}$. Fig.(\ref{wstar_vs_f}) shows the 
 dependence of $\omega^*$ with the average force $\fm$, where we 
 observe the scaling:
 \be
 \label{2b}
 \omega^*\sim \langle f\rangle^{1/2},
 \ee
 which holds well from the glass transition toward our densest packing, up to $\langle f\rangle =10^4$ for our system size.
 This scaling behaviour  in the vibrational spectrum  can be deduced from Eqs.(\ref{01},\ref{12}) if marginal stability is assumed throughout the glass phase.

Both observations (i) and (ii) bring further support on the marginal stability of the meta-stable states, previously inferred from the microscopic structure.

\subsection{Microscopic dynamics and normal modes}

To study the microscopic dynamics, we  project the dynamics on the normal modes and define
 for each frequency $\omega$:
\ba
C_{\omega} (t) =  \langle~ \langle \delta {\bf R}(t+t_w) | \delta {\bf R}^{\omega} \rangle ~.~
  \langle \delta  {\bf R}(t_w) |  \delta {\bf R}^{\omega}  \rangle   ~\rangle_{t_w},  
\label{Cw_alpha}
\ea 
where $| \delta {\bf R}(t) \rangle\equiv |{\bf R}(t)\rangle - |{\bf R}_{eq}\rangle$ is the displacement field around the 
 configuration corresponding to the average particles position,
and where the average is made on all time segments $[t_w,t_w+t]$  entirely
 included in a meta-stable state.

\begin{figure}                                
\rotatebox{-90}{\resizebox{6.50cm}{!}{\includegraphics{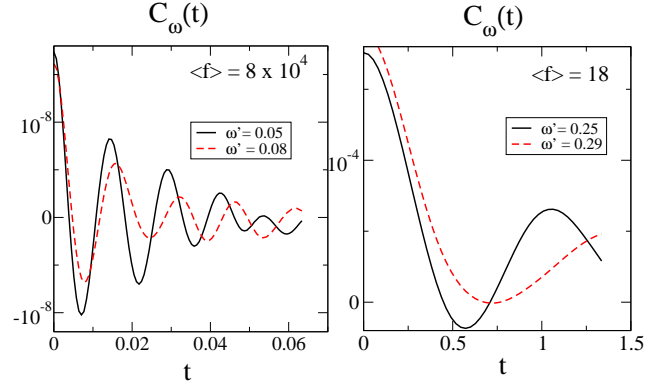}}}
 \caption{Examples of $C_{\omega}(t)/C_\omega(0)$  for low-frequency modes for two different average contact force
   $\fm$,  deep in the glass phase (left) and in the
   super-cooled liquid phase (right).}
   \label{correl_modes}
\end{figure}

If the projections of the dynamics  were made on longitudinal plane waves rather
than on normal modes, $C_{\omega} (t)$ would simply correspond to the
de-correlation of the density fluctuations
at some wave vector, which can be probed in scattering
experiments. 
Examples  of $C_{\omega}(t)$ for some  low-frequency  modes are presented in
Fig.(\ref{correl_modes}) at  two different  pressures, deep in
the glass phase and in the super-cooled liquid. We observe damped oscillations  for most of the spectrum.

From $C_{\omega}(t)$, the amplitude $A(\omega)$ and the 
characteristic time $\tau(\omega)$ of the oscillations of a mode are
 readily extracted.  The average square amplitude of the normal mode 
follows $\langle A^2(\omega)\rangle =C_{\omega}(0)$. We define the 
relaxation time scale $\tau(\omega)$ as the time at which $C_{\omega}(t)$ 
has decayed by some fraction $s$:  $C_\omega(\tau(\omega)) = s C_{\omega}(0)$.
 We have tried various definitions $s=0.3 ; 0.5 ; 0.9$ and found similar 
scaling for the dependence of $\tau(\omega)$ with $\omega$.
In what follows we present the data with $s=0.9$ where our statistic is more accurate. 

\begin{figure}                                
   \rotatebox{-90}{\resizebox{5.9cm}{!}{\includegraphics{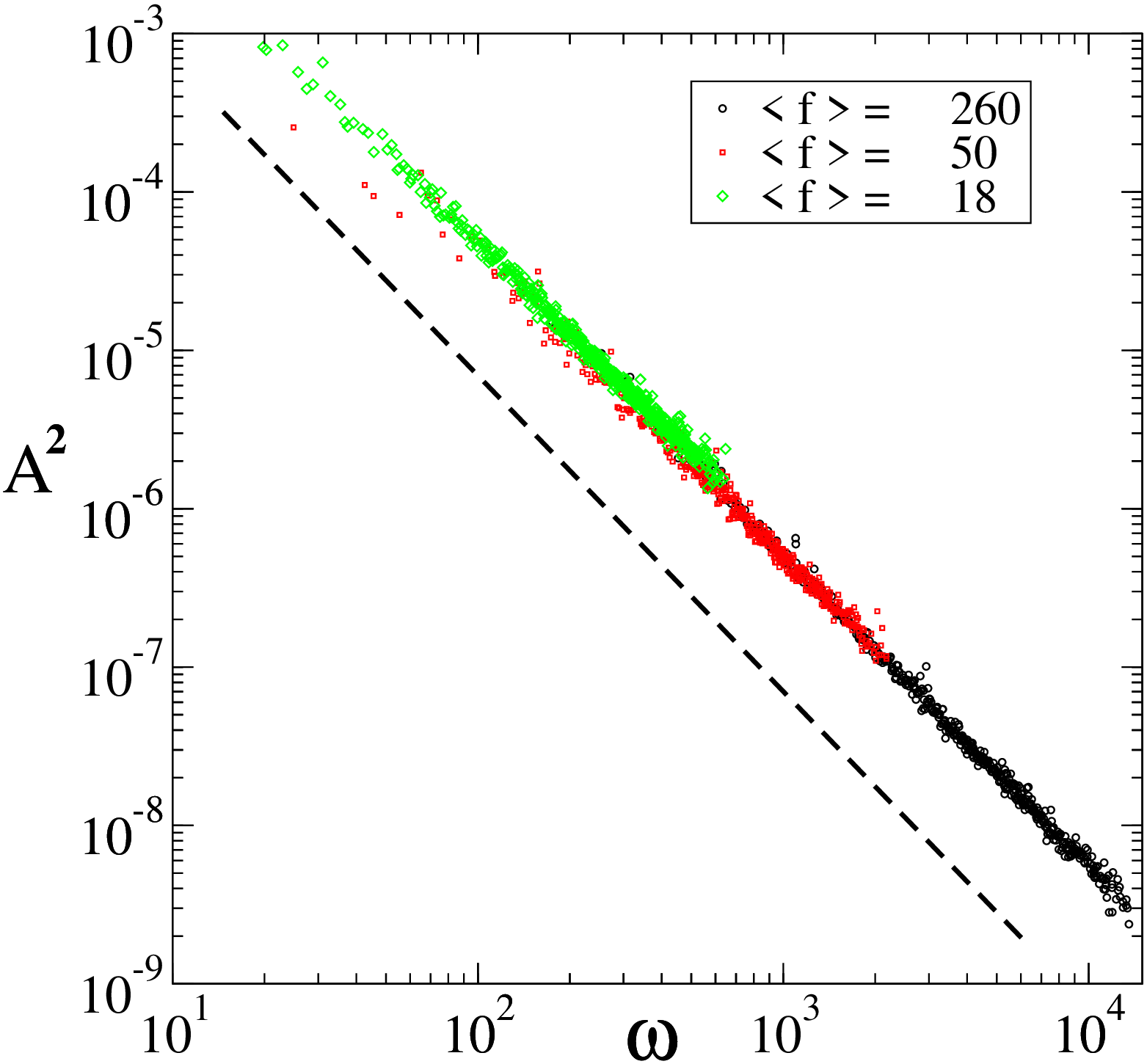}}}
   \vspace{-0.2cm}
   \rotatebox{-90}{\resizebox{5.9cm}{!}{\includegraphics{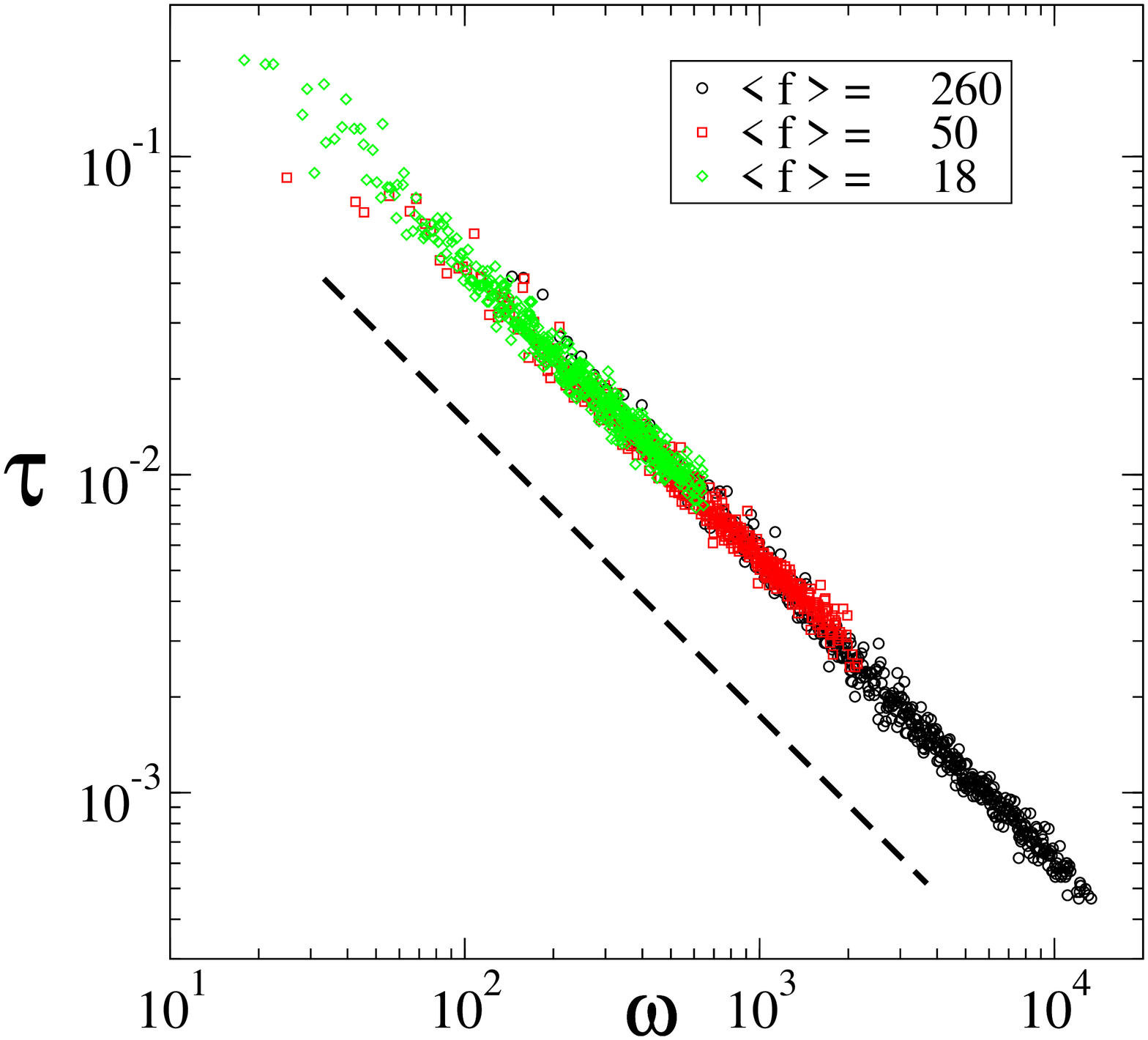}}}
   \caption{(a) Average squared amplitude of the modes 
     $\langle A^2(\omega)\rangle $ {\it vs} $\omega$ at 
     various packing fractions in a system of 256 particles, 
     both in the glass phase ($\fm=260$ and $\fm=50$) or in the 
     super-cooled liquid ($\fm=18$).
     Each point corresponds to one mode. The dashed line corresponds to the 
     fit $\langle A^2(\omega)\rangle\sim 1/\omega^2$.
     (b) Relaxation time $\tau(\omega)$ of each mode {\it vs} $\omega$ for the 
     same packing fraction.
     The slashed line corresponds to the relation $\tau(\omega) \sim 1/\omega$.}
    \label{A_vs_w}
\end{figure}

The dependence of these quantities with frequency are respectively
 shown in  Fig.(\ref{A_vs_w}) for three 
meta-stable states at different pressure.
 Two configurations are in the glass phase and one in the super-cooled 
liquid phase. 
In all cases, these quantities were computed for each mode of the spectrum.
We observe that the modes display weakly damped oscillations, whose 
amplitude and period follow:
 \ba
 \langle A^2(\omega)\rangle\sim \frac{1}{\omega^2}, \\
\label{rtt1}
 \tau(\omega) \sim \frac{1}{\omega}.
 \label{rtt}
 \ea
These results hold true even for the low-frequency part of the
spectrum, although  more scattering is found there 
\footnote[3]{As we observed before, sometimes one or a few unstable
 modes are observed. In this case the values of $\langle A^2 \rangle$
 and $\tau(\omega)$ are found to
 be of the order of those of the lowest-frequency stable modes.}.
 
 As a consequence, our computation of $D(\omega)$ gives a rather faithful distribution 
of relaxation time scales of the microscopic dynamics, supporting further the approximation we used to compute the free energy in Eq.(\ref{11}), a priori strictly valid only at infinite pressure. This allows us to identify the 
peak apparent in the inset of Fig.(\ref{Dw_examples}) as the Boson Peak, which appears
 as a similar hump in Raman or neutron spectra in molecular liquids
\cite{tao, angell, nakayama}. 
 Near the glass transition, this peak  appears at a frequency significantly smaller
 than $\omega^*$,  as shown by the inset of the Fig.(\ref{Dw_examples}) .

\subsection{Mean squared displacement}

In this section we use $D(\omega)$  to compute the mean square displacement around 
an equilibrium position inside a meta-stable state when $\fm$ is varied.
  This quantity is directly
 related to the Debye-Waller factor accessible empirically with scattering experiments.

We define $\delta {\vec R_i}= {\vec R_i}-{\vec R_i^{eq}}$, where ${\vec R_i^{eq}}$ 
is the average position of particles $i$ in a given meta-stable state as defined in 
Eq.(\ref{14}). Assuming that the dynamics 
of different modes is independent, the fluctuations of particles positions
 $\langle {\delta \vec R_i}^2\rangle$ can be written as a sum of the fluctuation of all modes:
\ba
\label{222}
\langle {\delta \vec R_i}^2\rangle= \sum_\omega \langle A^2(\omega) \rangle \langle {\delta \vec R_i(\omega)}^2\rangle 
\ea
where $A^2(\omega)$ is the average square amplitude of the amplitude
of the mode $\omega$ and ${\delta\vec R_i(\omega)}$ is the displacement of particle $i$ for the mode $\omega$.
We then average on all particles and define $\langle\delta \vec R^2 \rangle =  1/N~\sum_i \langle {\delta \vec R_i}^2\rangle$ where $N$ is the system size. 
Using the modes normalization  $\langle {\delta \vec R_i(\omega)}^2\rangle_i = 1/N$ 
and applying  Eqs.(\ref{rtt1}) lead to:
\be
\label{22}
 \langle {\delta \vec R}^2\rangle\sim \int_0 \frac{D(\omega)}{\omega^2} d\omega\geq \int_{\omega^*}\frac{D(\omega)}{\omega^2} d\omega.
\ee
The inequality accounts for the  modes with frequency between 
$\omega=0$ and $\omega^*$ that we have neglected. Accounting for those modes would not change our conclusion
as long as the soft modes density grows sub-linearly at low frequency. 
As can be checked for the square lattice, $D(\omega)$ reaches a typical value $1/\sqrt k$ ($\sim 1/\fm$ for hard spheres) for
 $\omega\geq \omega^*$. This is more generally true for amorphous packing, as proven in \cite{matthieu1}. Using this fact,  the last
 integral is dominated by the lowest bound and one gets:
\be
\label{23}
\langle {\delta \vec R}^2\rangle \geq \frac{D(\omega^*)}{\omega^*}\sim \fm^{-3/2}\sim h^{3/2}
\ee
which holds in any dimension $d\geq 2$ (with corrections of order $h^2 \log N$ for $d=2$ due to plane waves).  
We have used the scaling of the frequency scale $\omega^*$ confirmed in 
Fig.(\ref{wstar_vs_f}).
In crystals, the fluctuations around a particle position is of the order
 of the inter-particle gap $h$:  
$\langle \delta{\vec R}^2\rangle \sim h^{2} $ 
(with $\log N$ corrections in two dimensions).
Eq.(\ref{23}) shows that, near maximum packing,  the 
amplitude of particles motions is infinitely  smaller in the crystal
than in the glass. Because of  the marginal stability of the glass, 
these fluctuations have an  anomalous scaling with the packing fraction.

\begin{figure}
   \begin{center}                             
     \rotatebox{-90}{\resizebox{6.0cm}{!}{\includegraphics{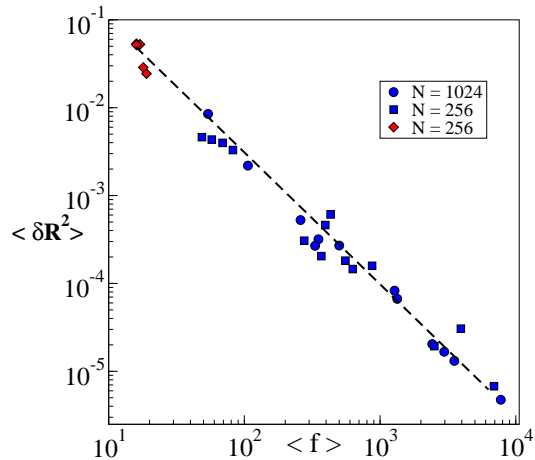}}}
     \caption{Mean square displacement $\langle \delta \vec R^2 \rangle$ versus
       average contact force $\fm$ for $N=1024$ (circles) and $N=256$
       (squares) particles. Diamonds correspond to the
       super-cooled liquid phase and were computed for a system with $N=256$ 
       particles.
       Slashed line corresponds to the best fit agreeing with our prediction 
       $\langle \delta \vec R_i^2 \rangle \sim \fm^{-3/2}$.}
     \label{sqrt_r2} 
   \end{center}                             
\end{figure}

 To check numerically this prediction, we consider various meta-stables states.
 In each of them, we measure  $\vec R_i^{eq}$ and the mean square displacement  
around the equilibrium position:
$\langle \delta \vec R^2 \rangle = \langle 1/N~\sum_i \delta \vec
R_i^2(t)  \rangle_{t_1} $, where the average is made on the time interval $t_1$.
 Fig.(\ref{sqrt_r2}) show this quantity for various packing fraction. Our numerical 
result agrees well with our prediction
$\langle \delta \vec R^2 \rangle \sim \fm^{-3/2}  \sim h^{3/2}$ throughout the glass phase.

\section{$\alpha$-relaxation}

One long-lasting challenge in our understanding   of the glass transition is the elaboration of a spatial description of activated events,
the rare and sudden rearrangements of particles corresponding to jumps between meta-stable states. These events are collective rearrangements of particles,
but the cause and the nature of this collective aspect is unknown. Our observation that the glass structure is marginally stable suggests that the softest, barely stable modes 
may play a key role in the activated events that relax the structures. In what follows we investigate this possibility by projecting the sudden rearrangements on the normal modes of the free energy. 

\subsection{Aging}
\label{glass}

During aging, sudden rearrangements, or ``earthquakes",
appear as drops in the self scattering function, 
 see Fig.(\ref{Cq_examples}-a).
 Such earthquakes correspond to  collective motions of a large number of particles,
 and  have been  observed in various other aging systems, such as colloidal paste or
 laponite \cite{duri}, and in Lennard-Jones simulations \cite{lapo, kb, heuer}.  
 Even for our largest numerical box of $N=1024$ particles, 
 deep in the glass phase these events generally  span the entire
 system. Examples of earthquakes  in real space   are shown in  
  Fig.(\ref{RD_AVEC-GLASS}). 

\begin{figure}   
     \begin{center}
       \rotatebox{-90}{\resizebox{6.0cm}{!}{\includegraphics{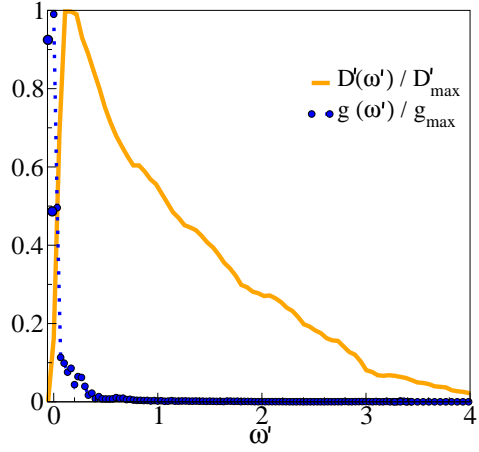}}}
       \caption{Straight curve: $D'(\omega')/D'_{max}$  {\it vs.}   $\omega/ \langle
         f\rangle$. 
         Both  $D(\omega)$ and $\fm$
         are computed in  the meta-stable state prior to the earthquake
         shown in the Fig.(\ref{Cq_examples}-a).
         Dotted curve: $g(\omega)$ as defined in the text re-normalized by its maximum value 
         $g(\omega') / g_{max}$  {\it vs.} $\omega/\langle f\rangle$.
         This figure has been originally published in the reference ``Heterogeneous dynamics, marginal stability and soft modes in hard sphere glasses'', J. Stat. Mech., L08003, (2007), by C. Brito and M. Wyart.}
   \label{Dw_proj}
 \end{center}
\end{figure}

To analyze these displacement fields, we measure 
the average particle positions and the contact network in the meta-stable state prior to the
earthquake, and compute the normal modes of the free energy.  
We also compute the earthquake  displacement $|\delta {\bf R}^e\rangle$ defined as a
 difference between the average particles position in two successive
 meta-stable states $l$ and $m$: 
 $|\delta {\bf R}^e\rangle\equiv |{\bf R}^m\rangle- |{\bf R}^l\rangle$.
 We then project $|\delta {\bf R}^e\rangle$ on the  normal modes and compute
 $c_\omega\equiv\langle \delta {\bf R}^e|\delta {\bf R}^\omega\rangle/ \langle \delta {\bf R}^e|
 \delta {\bf R}^e\rangle$. 
 The $c_\omega$'s  satisfy $\sum_\omega c_\omega{}^2=1$ since the normal
 modes form a unitary basis. 
 To study how the contribution of the modes  depends on 
 frequency, we define:
 \be
 g(\omega)= \langle c_\omega{}^2\rangle
 \ee
  where the average is made on a small segment of frequencies $ [\omega,
 \omega+d\omega]$. Fig.(\ref{Dw_proj}-a)  shows $g(\omega)$ for the earthquake
 shown in Fig.(\ref{Cq_examples}-a). The average contribution of
   the modes decreases very rapidly with increasing frequency, and  
   most of the displacement projects on the excess-modes present near zero-frequency. 
   This supports that  the free energy barrier crossed by the system during a rearrangement lies in the direction of the softest degrees of freedom.

\begin{figure}[h]
   \rotatebox{-90}{\resizebox{6.0cm}{!}{\includegraphics{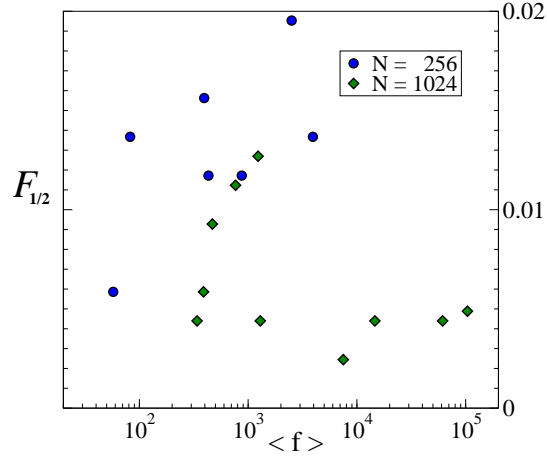}}}
   \caption{$F_{1/2}$ {\it vs.}  $\langle f \rangle$ for  
    $N=256$ (circles) and $N=1024$ (diamonds) particles.
     This figure has been originally published in the reference ``Heterogeneous dynamics, marginal stability and soft modes in hard sphere glasses'', J. Stat. Mech., L08003, (2007), by C. Brito and M. Wyart.}  
  \label{Fk_alpha2}
\end{figure}

\begin{figure}
  \hspace{-1.5cm}
  \rotatebox{-90}{\resizebox{9.0cm}{!}{\includegraphics{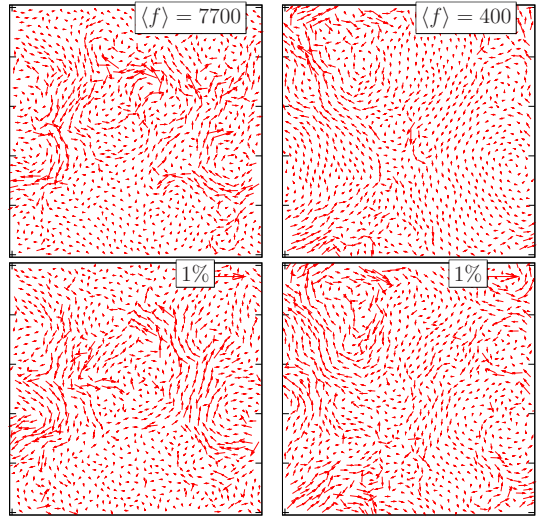}}}
  \caption{Above: two examples of earthquakes in the glass
    phase for different average contact force $\langle f \rangle$ for $N=1024$ particles.
    Displacements  were multiplied by four for visibility.
    Below:  projection of earthquakes    on the  $1\%$ of the normal   modes that  contribute most to them.}
  \label{RD_AVEC-GLASS}
\end{figure}

To make this observation systematic, introduce the label $i$ to rank the
 $c$'s  by decreasing order: $c_1>c_2...>c_{2N}$. We then defined a $k_{1/2}$ such that:
\be
\sum_{i=1}^{k_{1/2}} c_i^2 \equiv 1/2
\ee
 Physically, $k_{1/2}$ is the 
minimum number of modes necessary to reconstruct $50\%$ of the displacements relative to 
the earthquake.
Fig.(\ref{Fk_alpha2}) shows $F_{1/2}\equiv k_{1/2}/(2N)$ for the 17 cracks 
studied and indicates that  $0.2\%<F_{1/2}< 2\%$  for all the events studied
 throughout  the glass phase.
We thus systematically observe that the extended earthquakes correspond  
to the relaxation of a small number  of degrees of freedom, of the order of $1\%$
 of the modes of the system.

 In Fig.(\ref{RD_AVEC-GLASS}) we illustrate the spatial consequence of our analysis. Two examples of earthquakes
at  different  packing fractions are compared with the linear superposition of the $1\%$ of
the modes that contribute most to them. The similarity is striking: the complexity of the structural relaxation is indeed contained in the soft degrees of freedom of the system,
along which yielding occurs. Thus, in this regime only a small fraction of the degrees of freedom of the system participates in the relaxation of the structure.

\subsection{Structural Relaxation in the equilibrated super-cooled  liquid}
\label{SCL}

We equilibrate the system for a range of density $0.77\leq\phi\leq0.786$.
Also in  this regime, the dynamics  is heterogeneous in space and 
in time and sudden 
rearrangements still occur on time scales of the 
order of $\tau$ \cite{kob}.
An example of this rearrangement, that can be  identified as a drop in the 
self-scattering function, is shown  in the Fig.(\ref{Cq_examples}-b).
In  real space, this displacement corresponds to a collective event, 
as one can observe in the examples of the Fig.(\ref{rearrangement_N64_SL}-left) and
 Fig.(\ref{RD_AVEC-SL}-above).
To study these events in an equilibrated super-cooled liquid, we  extend the  
procedure used in the aging regime: we identify the 
 meta-stable states visited by the dynamics and compute their averaged configuration.
 We then define the normal  modes in the meta-stable state and the displacement field 
 corresponding 
to the relaxation events. For each relaxation event, we compute  $F_{1/2}$.

\begin{figure}
  \begin{center}
    \rotatebox{-90}{\resizebox{6.0cm}{!}{\includegraphics{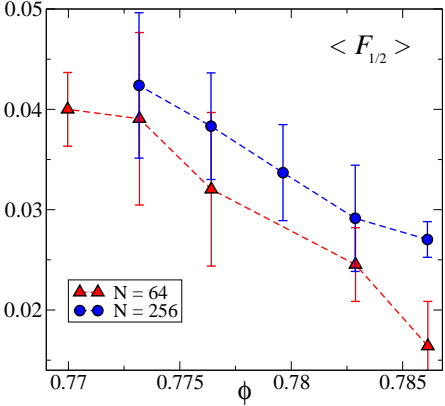}}}
    \caption{ $\langle F_{1/2} \rangle$ {\it vs} $\phi$ for two different system sizes.
      This figure has been originally published in the reference ``Heterogeneous dynamics, marginal stability and soft modes in hard sphere glasses'', J. Stat. Mech., L08003, (2007), by C. Brito and M. Wyart.}       
    \label{F_vs_phi_256_64}
    \end{center}
\end{figure}

\begin{figure}
  \begin{center}
    \rotatebox{-90}{\resizebox{8.0cm}{!}{\includegraphics{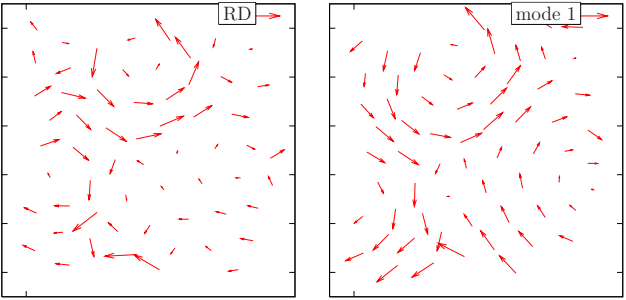}}}
    \vspace{-1.0cm}
    \caption{Left:  displacement field for a system with $N=64$ particles corresponding to a relaxation event. Arrows were multiplied by $1.2$. Right: normal mode that contains  $80\%$ of the projection of the real displacement field. This normal mode has  the lowest frequency of the spectrum.
      This figure has been originally published in the reference ``Heterogeneous dynamics, marginal stability and soft modes in hard sphere glasses'', J. Stat. Mech., L08003, (2007), by C. Brito and M. Wyart.}       
    \label{rearrangement_N64_SL}
    \end{center}
\end{figure}

We start with a system with $N=64$ particles.
For each packing fraction, $F_{1/2}$ is computed for six  relaxation events.
Then  this quantity is averaged on  all events. The result of these
 $\langle F_{1/2}\rangle$ are shown in Fig.(\ref{F_vs_phi_256_64})  as a 
function of the packing fraction. 
We find that  $\langle  F_{1/2}\rangle\leq 5\%$ for all $\phi$ studied,
supporting that only a small fraction of the low-frequency modes contribute to the 
structural relaxation events also in this region of the super-cooled liquid.
This fraction decays  significantly as $\phi$ get closer to $\phi_0$, suggesting 
 a rarefaction of the number of directions  along which the system
can  yield near the glass transition.
Fig.(\ref{rearrangement_N64_SL}) exemplifies this conclusion: 
for this particular case, the relaxation event shown on the left  projects almost entirely on 
  one normal mode, shown on the right in this same figure.
This mode turns out to be the lowest-frequency  normal mode of the free energy.

\begin{figure}
  \begin{center}
     \rotatebox{-90}{\resizebox{6.0cm}{!}{\includegraphics{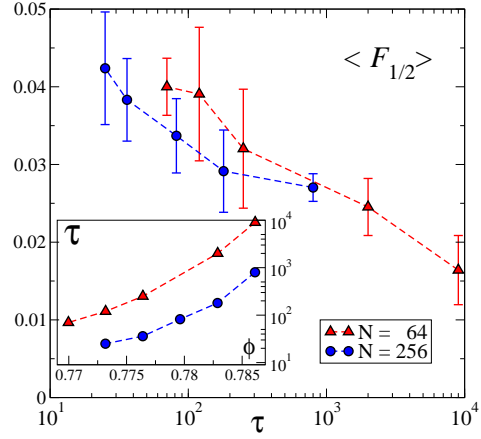}}}
     \caption{$\langle F_{1/2} \rangle$ {\it vs} $\tau$. 
       Inset: $\alpha$-relaxation time {\it vs} $\phi$. System sizes are indicated in the legend.
       This figure has been originally published in the reference ``Heterogeneous dynamics, marginal stability and soft modes in hard sphere glasses'', J. Stat. Mech., L08003, (2007), by C. Brito and M. Wyart.}       
     \label{F_vs_tau_256_64}
    \end{center}
\end{figure}

\begin{figure*}
  \rotatebox{-90}{\resizebox{9.0cm}{!}{\includegraphics{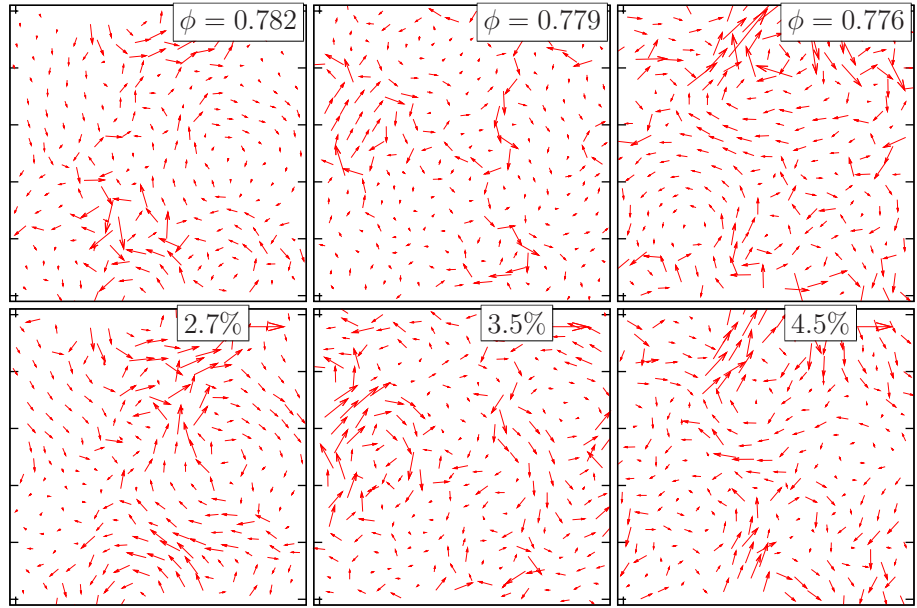}}}
  \caption{Above: Relaxation events in the super-cooled liquid for different $\phi$, indicated in the figure, and  $N=256$ particles.
    Displacement fields are rescaled by 4,  1.5 and 1.2 respectively for visibility.
    Below:  projection of the relaxation events on the normal  modes that  contribute 
    the most. The  fraction of the total number of modes used is indicated in each 
    figure, and corresponds to the fraction necessary to recover $50\%$ of the relaxation event. 
    As indicated in Fig.(\ref{F_vs_phi_256_64}), this fraction of modes tend to increase as $\phi$ decreases.}
  \label{RD_AVEC-SL}
\end{figure*}

To study  finite-size effects, we measure $F_{1/2}$ for twelve 
relaxation events at each of the five packing fraction using $N=256$ particles.
Results are shown in Fig.(\ref{F_vs_phi_256_64}).  Finite size effects are present, and $\langle  F_{1/2}\rangle$ appears to be roughly
 $0.5\%$ higher in the larger system for all packing fractions. Most of this difference in behavior is explained by the observation that the glass transition occurs at smaller packing fraction in the $N=64$ system, as previously observed \cite{yy}. The inset of Fig.(\ref{F_vs_tau_256_64}) shows that this is the case in our system as well.
If  $\langle F_{1/2} \rangle$ is plotted as a function of relaxation time, as in Fig.(\ref{F_vs_tau_256_64}),  the curves become similar for the two systems and 
$\langle  F_{1/2}\rangle$ is systematically smaller for a system with $N=256$ particles. 
Thus, even for larger systems,  collective rearrangements relaxing the system are ``soft": 
they project mostly into a small portion of the vibrational spectrum.
We verify spatially this observation in Fig.(\ref{RD_AVEC-SL}). Three examples of 
relaxation events at different  packing fractions are compared with the vector field 
which is a linear superposition of the modes that contribute most to them.
This relation between soft modes and relaxation has been recently supported by the observation  that regions where structural relaxation is likely to occur, said to have a high ``propensity", also display an abundance of soft modes \cite{Harrowell_NP}.

Interestingly, the soft modes that characterize marginally stable structures are in general rather extended objects,  as can be observed from the examples presented here.
Theoretically this is what one expects both in the square lattice, as justified by Eq.(\ref{001}), and in amorphous packing \cite{matthieu1,matthieu2}. In this light it does not seem surprising that
activated events are collective.

\section{A geometric interpretation of pre-vitrification}

\subsection{ Pre-vitrification}

\begin{figure}
  \begin{center}
    \rotatebox{0}{\resizebox{8.0cm}{!}{\includegraphics{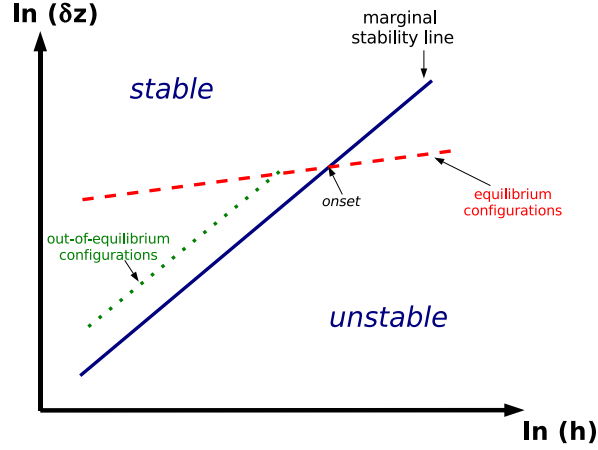}}}
    \caption{ Phase diagram for the stability of hard sphere configurations, in the coordination $\delta z$ {\it vs} average gap $h$ plane. The marginal stability line delimits stable and unstable configurations. The dashed  line correspond to equilibrium configurations for different $\phi$. As $\phi$ increases, $h$ decreases and the two lines eventually meet. This occurs at the onset packing fraction $\phi_{onset}$, where dynamics become activated. At larger $\phi$, viscosity increases sharply as configurations visited become more stable. For a finite quench rate the system eventually falls out of equilibrium. More stable and more coordinated regions cannot be reached dynamically, and as $\phi$ is increased further, the system lives close to the marginal stability region, as indicated in the dotted line. The location of the out-of-equilibrium trajectory depends on the quench rate. In the limit of very rapid quench, the out-of-equilibrium line approaches the marginal stability line.}
     \label{fig20} 
      \end{center}
\end{figure}

We have shown, both from its microscopic structure and microscopic dynamics, that the hard sphere glass lies close to marginal stability.
In this section we propose an explanation for this observation. This requires a single assumption, namely that the viscosity increases very rapidly when meta-stable states 
appear in the free energy landscape. In the logarithmic representation of the plane coordination {\it vs}  the typical gap between particles in contact $(\delta z, h)$,
there exists a line, corresponding to  the equality of  Eq.(\ref{12}), which separates a region where configurations are stable and unstable, as sketched in Fig.(\ref{fig20}).  At any packing fraction $\phi$,
equilibrium   configurations correspond to a point in the $(\delta z,
h)$ phase diagram.  As $\phi$ is varied equilibrium states draw a line
in this plane, represented by the dashed line (red online) in
Fig.(\ref{fig20}). At low $\phi$, gaps among particle are large and
configurations visited are unstable. As $\phi$ increases, the gaps
narrow, and configurations become eventually stable. This occurs at
some $\phi_{onset}$ when the equilibrium line crosses the  marginal stability line. 
At larger $\phi$, the viscosity increases sharply, so that on our numerical time scales equilibrium cannot be reached deep in the regions where meta-stable states are present.
As a consequence, the system falls out of equilibrium at some $\phi_0$
close but larger than $\phi_{onset}$. Configurations visited must
therefore lie close to the marginal stability line, as represented by the dotted line in Fig.(\ref{fig20}), since more stable, better-coordinated configurations cannot be reached dynamically. 

In this view, $\phi_{onset}$ corresponds to the onset temperature,
where activation sets in and the dynamics becomes intermittent.  When
intermittency appears, the $\alpha$-relaxation time scale $\tau$ is
still limited, and has increased roughly of one  order of magnitude
from the liquid state. This is consistent with the observation that
the configurations we probed in the super-cooled liquid, for which
$\tau$ is larger but still limited, are already stable: the free
energy expansion  has in general a positively-defined spectrum. Note
that at $\phi_{onset}$, $\delta z\neq 0$,
 and the characteristic length of the soft modes $l^*$ is finite. 
We think of those modes as involving a few tens of particles.

\subsection{Ideal glass and random close packing}

To put our work in a broader context it is useful to think about the phase diagram of Fig.\ref{fig20} with an extra dimension added.
For any configuration one can associate the packing fraction $\phi_c$ corresponding to the jammed packing that would be obtained after a rapid compression \cite{speedy, kurchan,zamponi}.
At equilibrium $\phi_c$ is an increasing function of pressure \cite{donev, berthier}. In the three-dimensional phase diagram $(h,\delta z,\phi_c)$, marginality is now represented by a surface. Its main feature,
the scaling relation between coordination and typical gaps expressed in Eq.(\ref{12}), holds irrespectively of the value of $\phi_c$ according to our theoretical analysis, in agreement with the data 
presented in Fig.(\ref{dz_vs_p}). Marginality and related properties are therefore adequately discussed in the more simple two-dimensional phase diagram presented in Fig.\ref{fig20}. 
Some other aspects of the dynamics and thermodynamics of hard spheres nevertheless benefit from introducing  the extra dimension $\phi_c$.

{\it Ideal glass:}
For molecular glasses the presence of an ideal glass transition where the configurational entropy  vanishes  at finite temperature has been proposed and is still debated \cite{wolynes,stillinger}. For hard spheres this view implies that the viscosity diverges at some finite $h>0$ \cite{kurchan,zamponi},  for which the equilibrium curve $\phi_c(h)$ reaches a constant value. 
Our work does not address the issue of the existence of an ideal glass, but it supports that if it exists, it is not responsible for the slow-down of the dynamics in the pre-vitrification region we can access empirically,
neither in the aging regime we could observe in the glass, since such a scenario would not explain the marginal stability of the microscopic structure we observe. Some authors have used  diverging fits of the relaxation time to argue in favor of the opposite view \cite{berthier}. Nevertheless, establishing the existence of an actual divergence from such fits  is questionable  even in molecular liquids \cite{dyre} where the number of decades of viscosities accessible is two to three times larger.

{\it Random Close Packing:} 
Empirically  it is observed that for various protocols of compression (such as pouring metallic balls in a container), the final packing fraction obtained is $\phi_c \approx 0.64$ for mono-disperse hard spheres.  This fact can be expressed as follows:  for each protocol one can associate a line $\delta z(h), \phi_c(h)$ characterizing the configurations visited during compression. Isostaticity implies $\delta z \rightarrow 0$ as $h\rightarrow 0$. Furthermore, for a wide class of protocols $\phi_c(h)\rightarrow 0.64$ as $h\rightarrow 0$. The explanation for this observation is debated \cite{kurchan}. An interesting hypothesis is that $\phi_c\approx 0.64$ corresponds to the limit reached by infinitely rapid compressions. If typical protocols are fast in comparison with the relevant time scales of the dynamics, they should generate packings with a similar packing fraction.

\section{Conclusion}

We conclude by a brief summary of our results and a few remarks. We have derived a geometric criterion for the stability of hard sphere configurations, and we have shown that in a hard sphere glass this bound is nearly saturated.
This supports that pre-vitrification occurs when the coordination is sufficiently large to counter-balance the destabilizing effect of the compression in the contacts. Nearly unstable modes are collective displacement fields, whose spatial extension  is governed by the coordination.  Once meta-stable states appear in the free energy, activation occurs mostly along a small fraction of these soft modes. This observation supports that these modes are the elementary objects to consider to describe activation. It also implies that structural relaxation must be cooperative, since the soft degrees of freedom are collective.

 We have observed that less and less modes participate to the structural relaxation as the packing fraction increases near $\phi_{onset}$.  It is tempting to speculate that, as the number of degrees of freedom allowing relaxation is reduced, the size of the cooperatively rearranging regions grows to eventually saturate at the extension of the softest modes $l^*$.   Nevertheless, a quantitative description of the relationship between soft modes and dynamical length scale remains to be built and tested. Other factors, such as  the possible presence of locally favored structure of high coordination or some other spatial heterogeneities of the structure, may also have to be taken into account.

Our analysis of the structural relaxation at equilibrium applies to the pre-vitrification region, corresponding to the intermediate viscosities that can be probed numerically. Similar time scales are accessible experimentally in shaken granular matter and colloidal glasses. Our work does not address the behavior of the equilibrium dynamics for very large packing fraction. As a consequence, it is possible that at much larger viscosities than those we probed, in particular near the glass transition in molecular liquids,  our observations on the nature of the structural relaxation may not apply, and soft modes may play no role in the dynamics. Nevertheless several observations support that soft modes and dynamics are related even for those large viscosities. In particular,  the intensity of  the boson peak, which indicates the presence of soft modes in the spectrum, strongly correlates with the glass fragility \cite{novikov}, a fact which is not captured by current theories of the glass transition. 

Finally, our geometric approach to pre-vitrification is consistent with
Goldstein \cite{goldstein} views, who proposed 40 years ago that the
glass transition is related to the emergence of meta-stable states in
the energy landscape.  Other more recent descriptions of the glass
transition, such as the mode coupling theory (MCT) of liquids
\cite{sjorgen}, make a similar prediction
\cite{wolynes,parisi,laloux,isabelle}, and it is interesting to compare this
approach to ours. Here we indicate several differences and analogies in the
respective conclusions: (i)  in MCT the predicted location of the elastic
instability corresponds to the onset packing fraction \cite{brumer}.  This is consistent with our observation that  when the dynamics becomes intermittent  ($\phi \geq \phi_{onset}$), the configurations visited 
have in general  a positively defined spectrum, displaying no unstable modes. This is also supported by previous results showing that the dynamics is dominated  by activation in this parameter range \cite{activation}. Nevertheless, MCT predicts diverging time scales \cite{sjorgen} and dynamical length scales \cite{bb} at the onset packing fraction, which are not observed. Fitting empirical data with such divergences \cite{hans} leads to a critical packing fraction $\phi_{MCT}$ significantly larger than $\phi_{onset}$. The interpretation of the extra fitting parameter  $\phi_{MCT}$, and its relation with the free energy landscape, is at present unclear. 
(ii) In MCT  the dynamics is computed via a resummation of a perturbation expansion in the non-linear interaction among modes, around a point where plane waves are un-coupled. In our case,  we use a variational argument \cite{matthieu1,matthieu2} to capture the properties of the linear soft modes whose stability is at play. This argument applies as well in covalent \cite{these_matt} and attractive glasses \cite{these_matt, ning2}. This leads to an estimation of a length scale $l^*$ characterizing soft modes, which depends on the coordination. This length scale has not yet found a correspondence within MCT, where non-trivial length scales appear from the dynamics \cite{bb} but  diverge near the elastic instability, unlike $l^*$.  (iii) In our approach, the key microscopic parameters determining the location of the transition are coordination and pressure. In MCT, an important parameter is the area under the first peak of the pair correlation function \cite{lapo}. These two views bear similarities, as the later quantity can be considered as a rough measure of coordination. It remains to be seen if MCT can capture the critical behavior of the marginal stability line observed at very large pressure. Exploring this possibility may clarify the physical meaning of the approximations that characterize MCT.

\begin{acknowledgments}
We thank L. G. Brunnet, G. Biroli, J-P. Bouchaud,  D. Fisher,  O. Hallatschek, S. Nagel, D. Reichman
and T. Witten for helpful discussion and L.Silbert for furnishing the initial 
jammed configurations. C. Brito was supported by CNPq and M. Wyart by the
Harvard Carrier Fellowship.
\end{acknowledgments}

\appendix

\section{Determination of the rattlers}

\label{det_rattlers}

Near maximum packing, a few percents of the particles are trapped in a
 large ``cages"  on which they apply a minuscule force in comparison 
to the typical contact forces in the system.   Such particles, called rattlers, do not 
participate to the rigidity of the structure: if  removed,
 stability is still  achieved. When we compute e.g. the coordination 
of the microscopic structure, we do not take  these particles into account. 

\begin{figure}[htbp]
  \begin{center}
    \rotatebox{-90}{\resizebox{6.0cm}{!}{\includegraphics{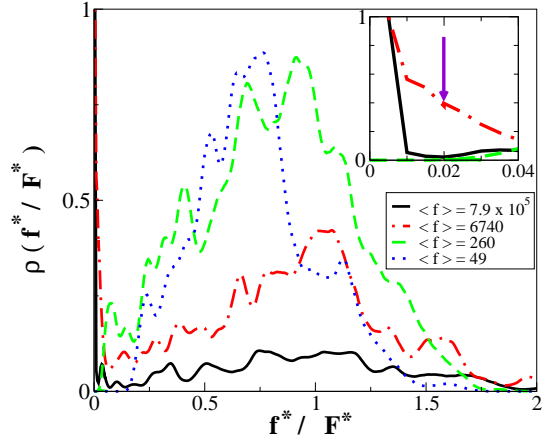}}}
    \caption{Histogram of the distribution of $f^*/F^*$ (see definition in the text) for various average force.}
    \label{forceDIST}
  \end{center}
\end{figure}

\begin{figure}[htbp]
  \begin{center}
    \rotatebox{-90}{\resizebox{6.0cm}{!}{\includegraphics{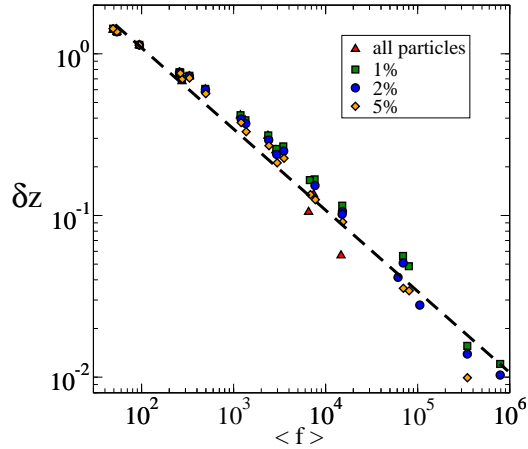}}}
    \caption{$\delta z$ vs $\langle f\rangle$ for 3 different criteria of  definition of rattlers as explained in the text. The legend ``all particles'' indicates that no rattlers are excluded for this measure. Dotted curve: fit of the relation $\delta z = A \langle f\rangle^{-1/2}$.}  
\label{diffCRIT}
\end{center}
\end{figure}

To identify rattlers we measure  the average number of 
shocks per  contact for each particle.
We compute how many shocks $n_{shoc}$ and how many 
contacts $n_c$  each particle has during the interval of time $t_1$ and
 define: $f^* = n_{shoc}/n_c$ if $n_{c}\geq 2$ and
 $f^* =0$ otherwise.
This quantity is normalized by the average number of shocks per
 contact that all particles have during $t_1$:
 $F^* = N_{shoc}/ N_{c}$, where $N_{shoc}$ in the 
 total  number of shocks and $N_{c}$ is the total number of contacts
in the system. 
 We then plot the distribution of $f^*/F^*$ for  different packing fractions,  
Fig.(\ref{forceDIST}).
 At large $\langle f \rangle$, we
observe the emergence of a peak near zero. When $\fm$ is intermediate,  
$\langle f \rangle = 5.2 \times 10^3$ and 
$\langle f \rangle = 9.2 \times 10^2$,  the peak vanishes.
This peak corresponds to the rattlers.  
In this work we consider that all particles for which
$f^*/F^* \leq 2\%$ are rattlers.
This criterion is represented by the arrow in the inset of the Fig.(\ref{forceDIST}).
To check the robustness of our results, we test if the relation between the excess of coordination 
$\delta z$ and the average force $\fm$ depends on this criterion.
We vary the threshold bellow which we  consider a particle as a  rattler
and plot in the Fig.(\ref{diffCRIT}) the comparison between 3 different criteria:
$f^*/F*  \leq 0.01  \leq 0.02 \leq 0.05$.
We observe that relation $\delta z = A_1 \fm^{-1/2}$ 
holds irrespectively of the criterion. It fails  when the rattlers are not removed 
of the analysis. In this case, for high values of $\langle f \rangle$  one finds $\delta z <0$. 

\section{Persistence of the normal modes in a meta-stable state}
\label{normal_modes_persistence}

\begin{figure*}
\centering
\begin{tabular}{cc}
\epsfig{file=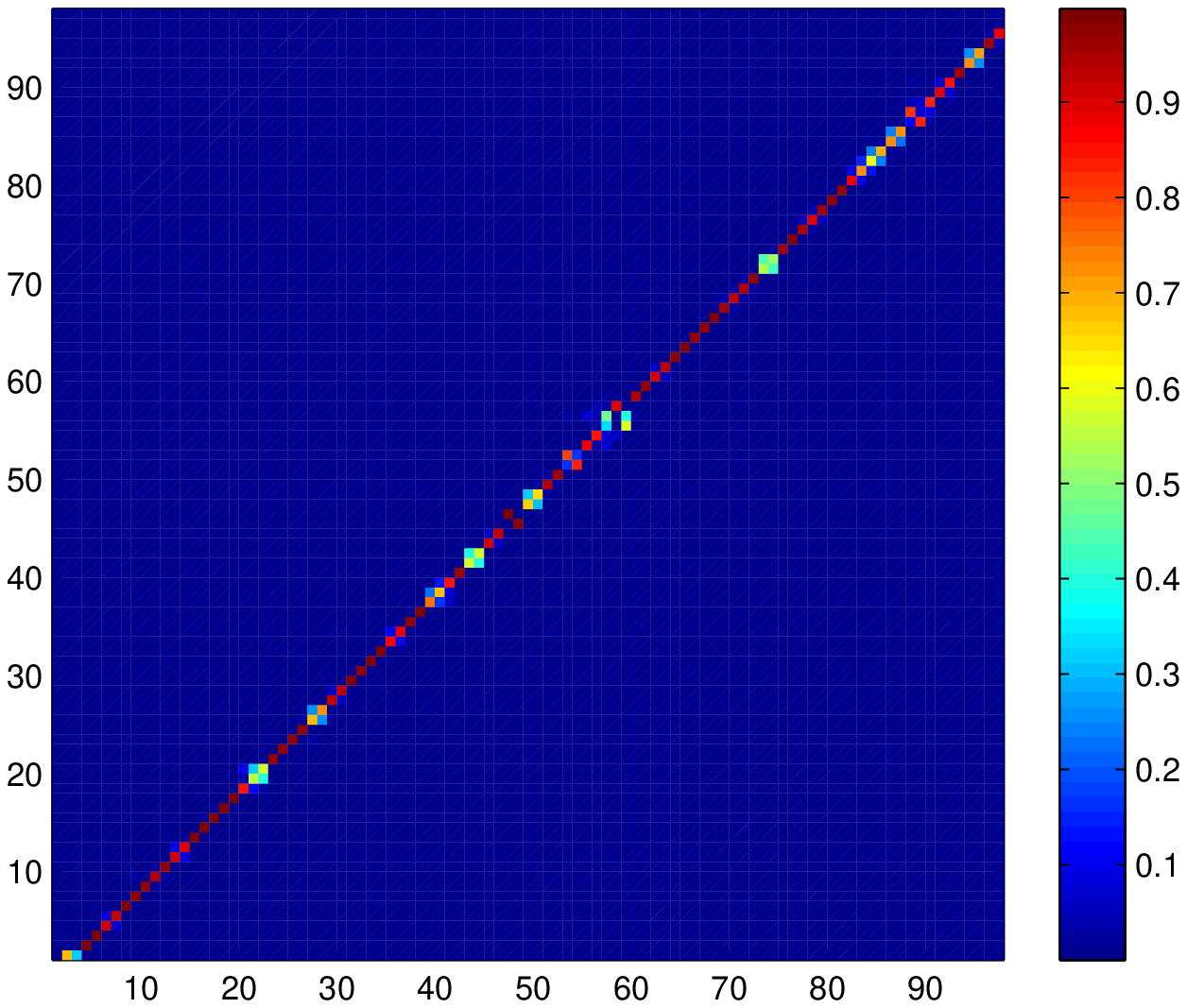,width=0.49\linewidth,clip=} &
\epsfig{file=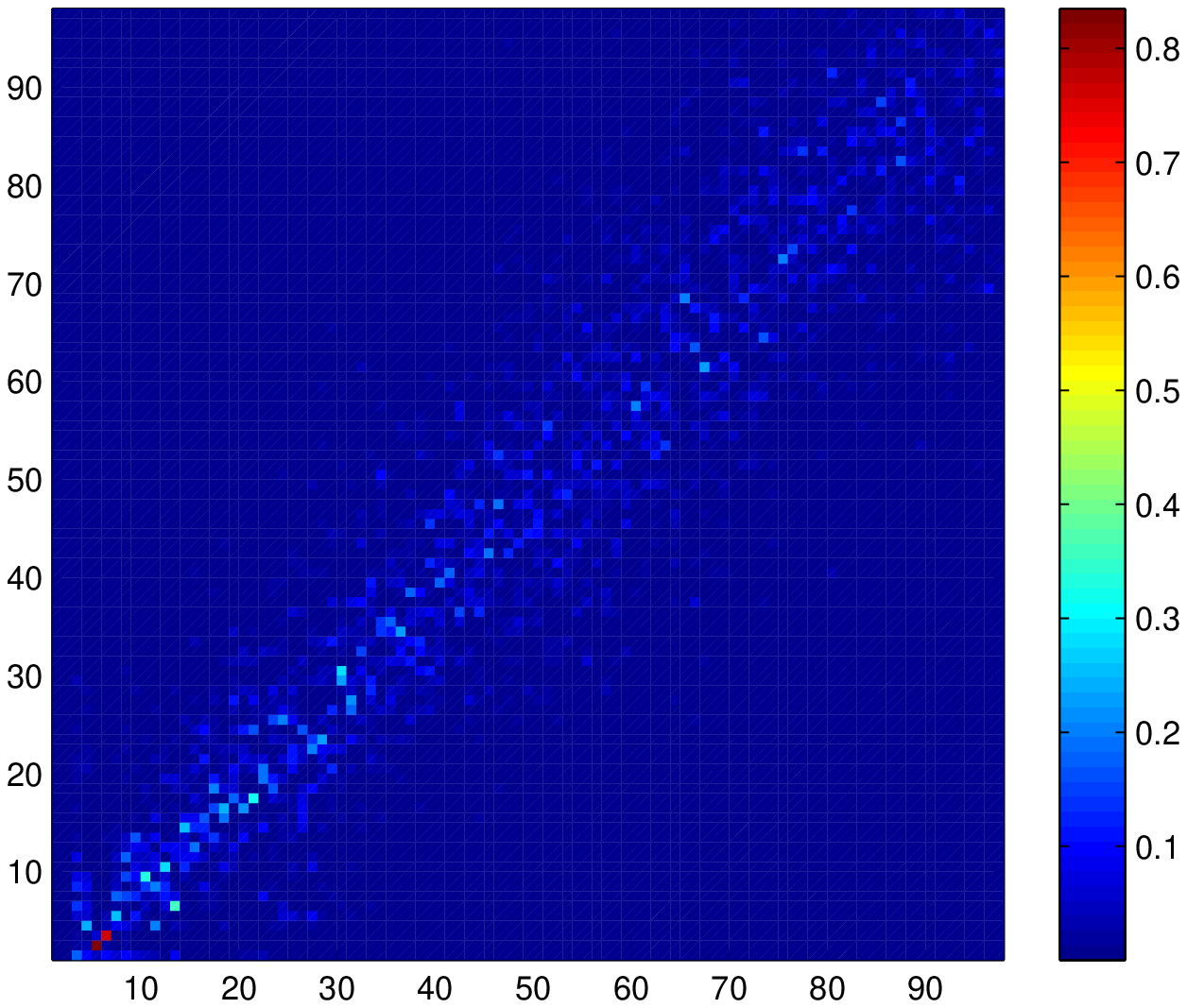,width=0.49\linewidth,clip=} 
\end{tabular}
\caption{Matrix of scalar product $C$ as defined in the text for the 100 lowest-frequency modes for $N=256$. 
The colorbar indicates the value of the scalar product. Time intervals lasted $t_1=500\tau_c$. 
Left:  $\fm=6700$.  The intervals $t_a$ and
$t_b$ are separeted by  $10^5 \tau_c$. 
Right: $\phi=0.782$. The intervals $t_a$ and
$t_b$ are separeted by  $10^4 \tau_c$, which is
of the order of the relaxation time.}
\label{scalar_product} 
\end{figure*}

Here we show that our numerical computation of the normal modes is robust to different choices of time intervals, as long as they lie in the same meta-stable state. 
To achieve that we compute the normal modes for two distinct, non-overlapping time-intervals. $| \delta {\bf R}_{t_a}^\omega\rangle$ denotes the normal modes of frequency $\omega$, computed on some time interval labeled $t_a$. 
We then compute the matrix of scalar product:
\be
C_{\omega,\omega'}= \langle \delta {\bf R}_{t_a}^\omega|\delta {\bf R}_{t_b}^{\omega '}\rangle.
\ee
If the two sets of normal modes are identical, $C$ is the identity matrix. In general this must not be exactly true, as shown in Fig.\ref{scalar_product},
since our protocol requires time-averaging and is therefore noisy to some extent, and also because some non-trivial dynamics may still occur within meta-stable states.
Our observations below show that those effects are small, even if the two time-intervals considered are separated by a time scale of the order of the life-time of meta-stable states. 
To quantify the difference between $C$ and the identity matrix, we follow the procedure we used before to compare a relaxation event to the normal modes of the structure.
We define $F_{1/2}(\omega)$ as the minimal fraction of normal modes computed on $t_b$ sufficient to represent $50\%$ of a normal model of frequency $\omega$ computed on $t_a$. 
We then define $\langle F_{1/2}\rangle$ as the average of $F_{1/2}(\omega)$ on the 20 lowest-frequency modes computed on $t_a$.  $\langle F_{1/2}\rangle$ is $1/2N$ if $C$ is the identity matrix,
and should be small if our procedure is robust to different choice of time-interval. This is indeed the case: in the super-cooled liquid ($\langle f\rangle=18$) we find $\langle F_{1/2}\rangle=0.4\%$,
which is small for all practical purposes discussed in this article. For this measure the time intervals lasted $t_1=500\tau_c$, and  the
time separation between $t_a$ and $t_b$ was $10^4 \tau_c$, which is
of the order of the relaxation time  $\tau\approx 3 \times 10^4 \tau_c$. For the glass ($\langle f\rangle=6700$) $F_{1/2}(\omega)$ is close to $1/2N$, as is obvious from Fig.\ref{scalar_product}.

\listoffigures

\end{spacing}

\end{document}